\documentclass[prc,showpacs,twocolumn,amsmath,amssymb,superscriaddress]{revtex4}
\usepackage{graphicx}
\usepackage{dcolumn}
\usepackage{bm}
\graphicspath{ {.} }

\begin{document}

\title{Microscopic description of quadrupole-octupole coupling in
 neutron-rich actinides and superheavy nuclei  
with the Gogny-D1M energy density functional}

\author{R. Rodr\'{\i}guez-Guzm\'an}

\email{raynerrobertorodriguez@gmail.com}

\affiliation{Department of Physics, Kuwait University, Kuwait }

\author{L.M. Robledo}

\affiliation{%
Center for Computational Simulation, Universidad Polit\'ecnica de 
Madrid, Campus Montegancedo, 28660 Boadilla del Monte, Madrid, Spain
}%

\affiliation{Departamento  de F\'{\i}sica Te\'orica and CIAFF, 
Universidad Aut\'onoma de Madrid, 28049-Madrid, Spain}

\email{luis.robledo@uam.es}

\date{\today}

\begin{abstract}
The interplay between quadrupole and octupole degrees of freedom is 
discussed in a series of neutron-rich actinides and superheavy nuclei 
with $92 \le$ Z $\le 110$ and  $186 \le$ N $\le 202$. In addition to 
the static Hartree-Fock-Bogoliubov approach, dynamical beyond-mean-field 
correlations are taken into account via both parity restoration and 
symmetry-conserving Generator Coordinate Method calculations based on 
the Gogny-D1M energy density functional. Physical properties such as  
correlation energies, negative-parity excitation energies as well as 
reduced transition probabilities $B(E1)$ and $B(E3)$ are discussed in 
detail. It is shown that, for the studied nuclei, the 
quadrupole-octupole coupling is  weak and to a large extent the 
properties of negative parity states can be reasonably well described 
in terms of the octupole degree of freedom alone.
\end{abstract}


\maketitle{}

%
%
%

\section{Introduction}
All over the nuclear chart, the majority of the spherical and/or 
quadrupole-deformed  ground states are  reflection-symmetric. However, 
in  regions with given proton and/or neutron numbers the spatial 
reflection symmetry is broken spontaneously and octupole-deformed 
ground states are favored energetically \cite{butler_96}. Octupole 
deformation  is also well known to affect the outer fission barriers of 
atomic nuclei and is the collective variable associated to cluster radioactivity (see, for example, 
\cite{Rayner-fission-1,Rayner-fission-5,Rep-Prog-Phys_SRobledo,cluster-rad.WardaRobledo}). 
The search for  signatures of octupole correlations has  remained  an 
active research field over the years 
\cite{Ahmad_93,butler_2016,butler_2015,Tandel_2013,Li_2014,Ahmad_2015,Bucher_2016,Bucher_146Ba_2017,Butler2020,Gaffney_2013,Chishti20}. 
Previous experiments have found evidence for octupole-deformed ground 
states in $^{144,146}$Ba \cite{Bucher_2016,Bucher_146Ba_2017} and 
$^{222,224}$Ra \cite{Butler2020,Gaffney_2013}  or measured the $E1$ 
strength in $^{228}$Th \cite{Chishti20}. Furthermore, a correlation 
between the Schiff moment \cite{Schiffmoment} and octupole deformation 
has been found  \cite{DobaczewskiSchiffmoment} suggesting that 
octupole-deformed nuclei might represent the best candidates for atomic 
electric dipole moment measurements. 

%

\begin{figure*}
\includegraphics[angle=-90,width=0.95\textwidth]{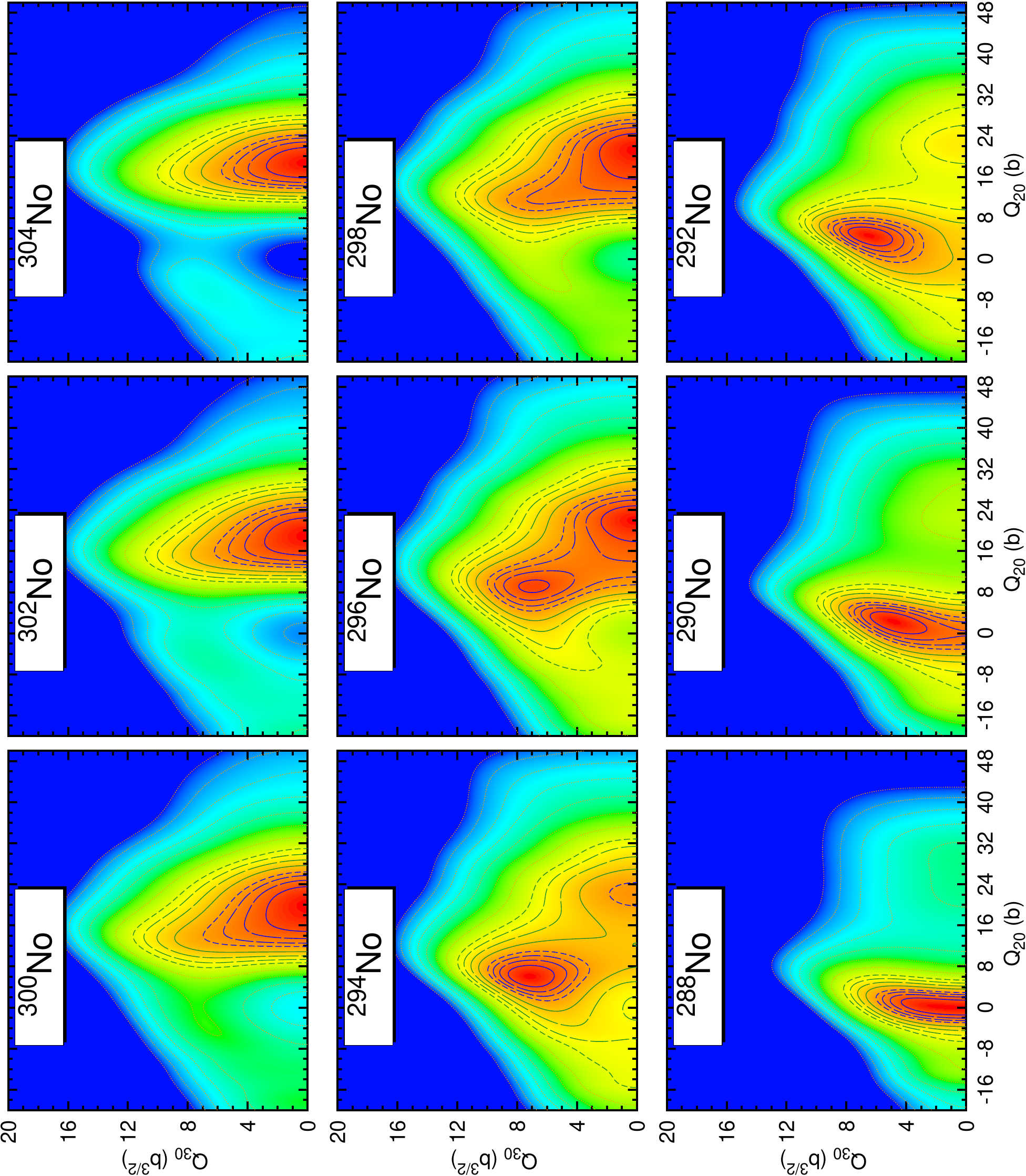}
\caption{(Color online) MFPESs computed with the Gogny-D1M EDF for the 
isotopes $^{288-304}$No. Dark blue contour lines 
extend  from 0.25 MeV  up to 
1 MeV above the ground state energy in steps of 0.25 MeV in the ascending 
sequence full, long-dashed, medium-dashed and short-dashed. Dark green
contour lines  extend  1.5 MeV up to 3 MeV above the ground state in steps
of 0.5 MeV with the same sequence of full, long-dashed, medium-dashed and
short-dashed as before. From there on orange dotted contour lines are drawn 
in steps of 1 MeV. The color code spans a range of 10 MeV with red
corresponding to the lowest energy and blue corresponding to an energy 10 MeV above. 
The intrinsic HFB energies are symmetric under 
the exchange $Q_{30} \rightarrow -Q_{30}$. For A = 294 a quadruple 
deformation $Q_{20}= 10 \textrm{b}$ is equivalent to $\beta_{2} = 0.141$ and an 
octupole deformation $Q_{30}= 1 \textrm{b}^{3/2}$ is equivalent to $\beta_{3} = 
0.021$. For more details, see the main text. 
}
\label{Q2Q3_MF_No} 
\end{figure*}

From a theoretical point of view, various models and approaches have 
already been employed to study the properties of octupole collectivity. 
Among them we can mention the  studies of octupole shapes  carried out using the 
macroscopic-microscopic (MM) approach 
\cite{moller_81,Gyurkovich-1981,naza_84,Nix-1995,Beng-2008} or the use 
of the Interacting Boson 
Model (IBM) \cite{Bing_2014,Nomura_Ba_RE_2018,Nomura_Th_RE_2018,Nomura_Rayner_IBM_2015}
with parameters determined using fermion-to-boson  mapping 
procedures starting from mean-field potential energy surfaces (MFPESs), obtained with 
relativistic and non-relativistic  energy density functionals (EDFs). 

Microscopic non-relativistic and relativistic approaches, both at the 
mean-field level and beyond, have been widely used to study octupole 
correlations 
\cite{mar83,bon86,bon91,hee94,erler-85,rob87,rob88,egi90,egi91,gar98,rob10,egi92,Long_2004,Tomas_GCM_parity_2016,Xu-2017,CDFT-beyondMF,JPG_2012_RoRay,Ebataba-2017,Xia_PRC_2017,Agbemava-Q3-2016,Agbemava-Q3-2017,Recent-Survey-Q3,Robledo-Bertsch-Q3-1,Robledo15}. 
Those microscopic studies include global surveys looking for octupole 
deformed mean field ground states in even-even nuclei  
\cite{Ebataba-2017,Xia_PRC_2017,Agbemava-Q3-2016,Agbemava-Q3-2017,Recent-Survey-Q3,Robledo-Bertsch-Q3-1,Robledo15}. 
In addition, properties of dynamic octupole correlations have been 
analyzed in large scale beyond-mean-field calculations carried out for 
even-even nuclei and using several parametrizations of the Gogny 
\cite{gogny} EDF \cite{Robledo-Bertsch-Q3-1,Robledo15}. The results of 
those calculations indicate that not only static octupole deformation 
but also dynamical beyond-mean-field octupole correlations have a 
sizeable impact on physical observables.

%

\begin{figure*}
\includegraphics[angle=-90,width=0.95\textwidth]{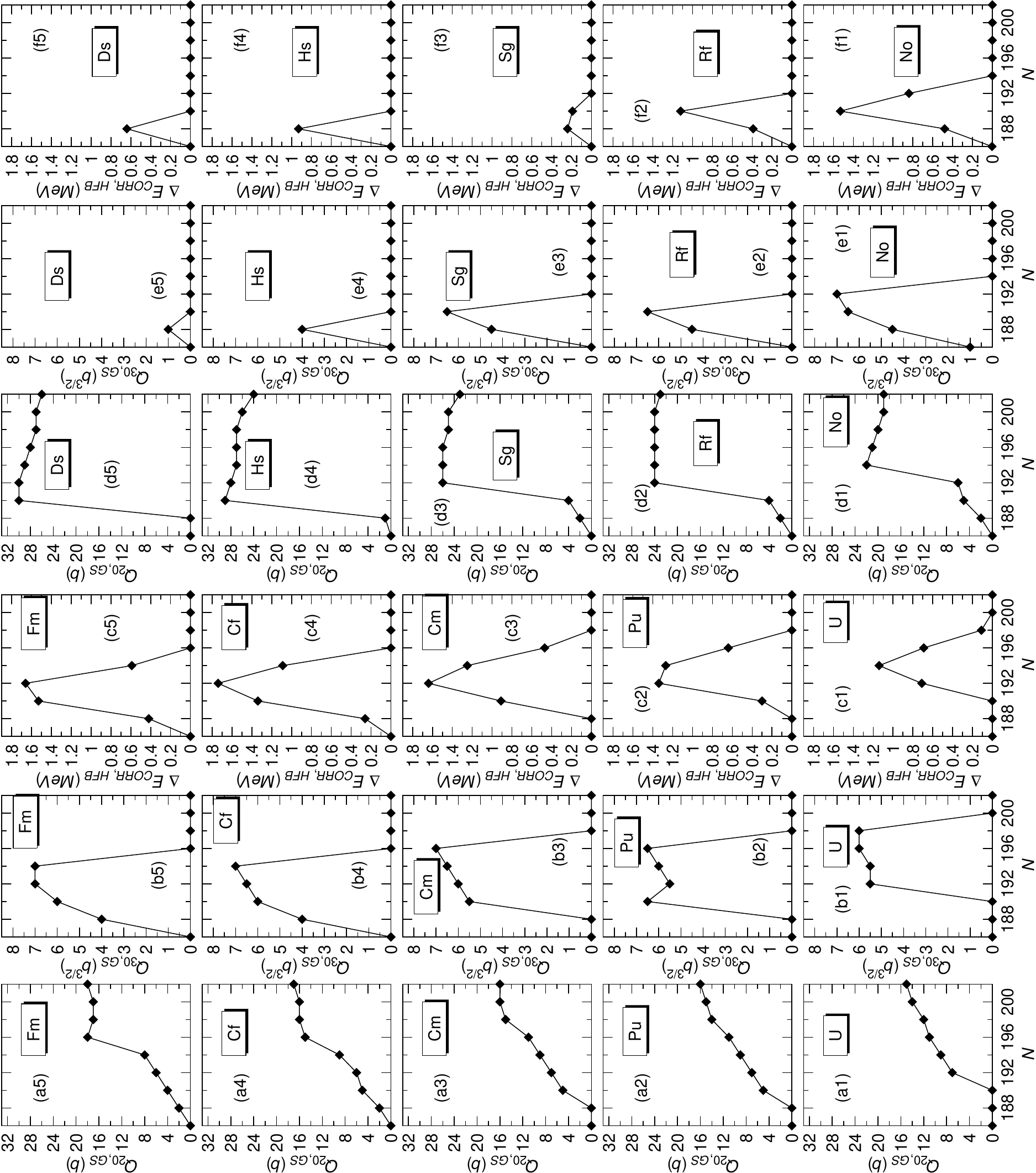}
\caption{The mean-field 
ground state quadrupole [panels (a1)-(a5) and (d1)-(d5)] and octupole 
[panels (b1)-(b5) and (e1)-(e5)] deformations as well as the octupole correlation 
energies $\Delta E_{CORR, HFB}$
[panels (c1)-(c5) and (f1)-(f5)]
Eq. (\ref{MFCorrEner}) are plotted, as functions of the 
neutron number, for the nuclei 
$^{278-294}$U, $^{280-296}$Pu, $^{282-298}$Cm, $^{284-300}$Cf,
$^{286-302}$Fm, $^{288-304}$No, $^{290-306}$Rf, $^{292-308}$Sg, $^{294-310}$Hs
and $^{296-312}$Ds. Results have been obtained with the Gogny-D1M EDF. 
}
\label{MF-2} 
\end{figure*}

The interplay between the two lowest multipole moments characterizing 
the nuclear shape, namely the quadrupole and octupole degrees of 
freedom has been studied in  Sm and Gd isotopes with neutron number 
$84 \le N \le 92$ \cite{Rayner_Q2Q3_GCM_2012} as well as in actinide 
nuclei around  $N \approx 134$ \cite{JPG_2020_UPuCmCf_RayRo}. 
Calculations have been carried out using the parameterizations D1S 
\cite{gogny-d1s}, D1M \cite{gogny-d1m} and D1M$^{*}$ 
\cite{gogny-D1MSTAR} of the Gogny-EDF. Both quadrupole and octupole 
constrains were considered simultaneously to build the MFPESs for the 
considered nuclei. Those MFPESs exhibited a  soft behavior along the 
octupole direction indicating, that dynamical beyond-mean-field  
effects should be taken into account. Those beyond-mean-field  effects 
were considered via both parity projection of the intrinsic states and 
symmetry-conserving quadrupole-octupole configuration mixing 
calculations, in the spirit of the two-dimensional (2D) Generator 
Coordinate Method (GCM) \cite{rs}. In addition to the systematic of 
the correlation energies, $1^{-}$ excitation energies, $B(E1)$ and 
$B(E3)$ transition probabilities, the results of 
Refs.~\cite{Rayner_Q2Q3_GCM_2012,JPG_2020_UPuCmCf_RayRo} indicate that 
2D-GCM zero-point quantum fluctuations lead to a dynamically enhanced 
octupolarity in the studied nuclei. The   2D-GCM framework  has also 
been applied to Rn, Ra and Th nuclei in 
Ref.~\cite{Robledo_2D-GCM_with_Butler}.

Octupole correlations in neutron-rich heavy and superheavy nuclei have 
been the subject of intense scrutiny in recent years 
\cite{Agbemava-Q3-2016,Agbemava-Q3-2017,Erler2012,Recent-Survey-Q3,Nix-1995,Jachi-2020, 
Fission-D1Mstarstar,Warda-Egido-2012}. Those nuclei will not be 
accessible with future Radioactive Beam Facilities (RBF). However,  
they represent the territories where the fate of the nucleosynthesis of 
heavy nuclei is determined and therefore  a better understanding of  their 
properties is required, for example, to improve the modeling 
of fission recycling in neutron star mergers 
\cite{nsmerger-1,nsmerger-2}. 

%

\begin{figure*}
\includegraphics[width=0.95\textwidth]{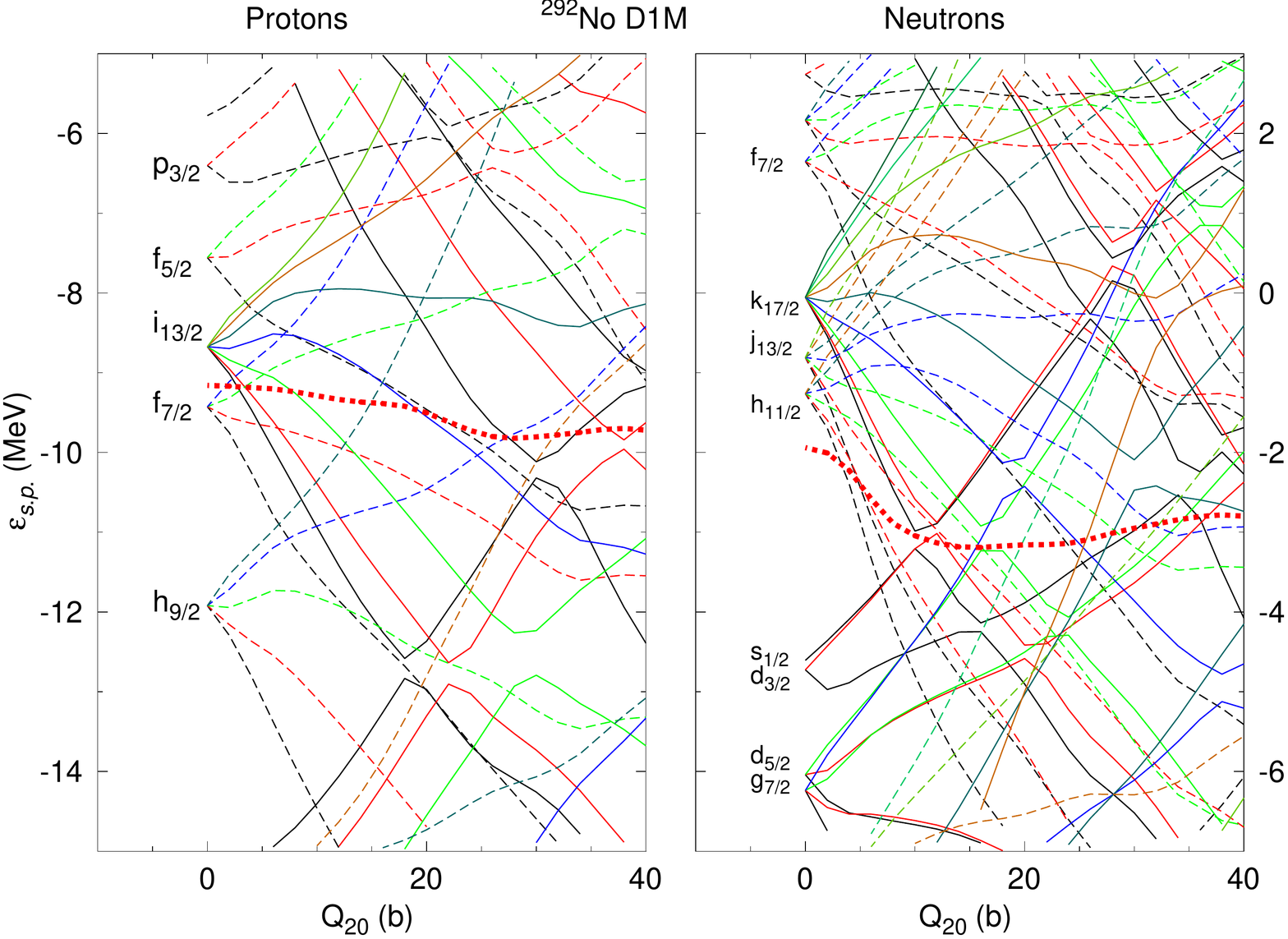}
\caption{(Color online) Single particle energies for protons (left panel) and neutrons
(right panel) as a function of the quadrupole moment and for reflection symmetric $Q_{30}=0$ shapes.
The spherical quantum numbers at $Q_{20}$ are also indicated in the plot.
The color code is associated with the $K$ value of the orbital (black, $K=1/2$;
red, $K=3/2$; green $K=5/2$; blue $K=7/2$; forest green $K=9/2$;  brown $K=11/2$, etc ).
Positive (negative) parity orbitals are plotted as full (dashed) curves.
The Fermi level is represented with a thick dotted red line. 
}
\label{MF-spe-RS} 
\end{figure*}

Among the calculations mentioned in the previous paragraph those based 
on the NL3$^{*}$, DD-ME2, DD-PC1 and PC-PK1 relativistic EDFs 
\cite{Agbemava-Q3-2016,Agbemava-Q3-2017}, have predicted an island of 
octupole-deformed nuclei in their ground state centered at (Z $\approx$ 
96, N $\approx$ 196).  An island of octupolarity has  also  been found  
with the SLy6 and SV-min zero range Skyrme-EDFs  but this time centered 
at (Z $\approx$ 100, N $\approx$ 190) \cite{Erler2012}. An inter-model 
comparison between the NL3$^{*}$, DD-ME2, DD-PC1 and PC-PK1 covariant 
EDFs and the UNEDF0, UNEDF1, UNEDF2, SLy4 and SV-min Skyrme-EDFs have 
been presented in Ref.~\cite{Recent-Survey-Q3} for Z $\le$ 110 and N 
$\le$ 210. It has been concluded that a region of octupole deformed 
ground state exists for 184 $<$ N $<$ 206. Furthermore, 
calculations within the MM framework  predicted an island of octupole 
deformation centered at  (Z $\approx$ 100, N $\approx$ 184) 
\cite{Nix-1995}. Recent MM large scale calculations for 98 $\le$ Z 
$\le$ 126 and 134 $\le$ N $\le$ 192 predicted octupole-deformed ground 
states for N $\ge$ 182 \cite{Jachi-2020}. Additionally, an
account  of the fission properties of superheavy nuclei 
with 100 $\le$ Z $\le$ 126 including very neutron-rich isotopes up to 
around 4 MeV from the two-neutron driplines, has predicted octupole 
instability for 186 $\le$ N $\le$ 194  using the Gogny-D1M$^{*}$ EDF \cite{Fission-D1Mstarstar}. All 
the aforementioned approaches agree on the existence of an island of 
octupolarity in neutron-rich actinides and low-Z (i.e., Z $\le$ 110) 
superheavy nuclei, in spite of the differences regarding its location 
and extension in the (Z,N)-plane. However, the predictions of different 
approaches differ for larger Z values 
\cite{Agbemava-Q3-2017,Erler2012,Nix-1995,Jachi-2020,Warda-Egido-2012,Fission-D1Mstarstar}.

Given the relevance of dynamical octupole correlations and/or symmetry 
restoration in the properties associated to the octupole degree of 
freedom pointed out in our previous studies discussed above we have 
decided to apply those techniques to the region of the nuclear chart 
including actinides and low-Z superheavies. In this work, we study the 
quadrupole-octupole coupling in neutron-rich even-even  nuclei with 
proton and neutron numbers $92 \le$ Z $\le 110$ and  $186 \le$ N $\le 
202$.  As in previous studies covering other regions of the nuclear chart
\cite{Rayner_Q2Q3_GCM_2012,JPG_2020_UPuCmCf_RayRo}, we consider three 
levels of approximation for each nucleus. First, the constrained 
Hartree-Fock-Bogoliubov (HFB) approach is used to obtain a set of mean 
field HFB wave functions, which are labeled by their intrinsic quadrupole and 
octupole moments. The energy associated with those HFB states is used 
to build a mean field potential energy surface (MFPES) which is a 
function of both the quadrupole and octupole moments. As discussed 
later on, those MFPESs often are rather soft along the octupole 
direction. Some of the studied neutron-rich nuclei display a pronounced 
competition, i.e., shape coexistence, between reflection-symmetric and 
reflection-asymmetric configurations. Moreover, in some cases the MFPES 
exhibits a transitional behavior along the quadrupole direction. 
Therefore, the HFB approximation can only be considered as a starting 
point and dynamical correlations stemming from the restoration of the 
broken parity symmetry (second level) and/or fluctuations in the collective quadrupole 
and octupole coordinates (third level) have to be taken into account.

The results discussed in this paper have been obtained with the finite 
range and density dependent Gogny-D1M EDF. Such a parametrization, 
specially tailored to better describe nuclear masses, has already 
provided a reasonable description of octupole properties 
\cite{JPG_2012_RoRay,Robledo-Bertsch-Q3-1,Rayner_Q2Q3_GCM_2012,JPG_2020_UPuCmCf_RayRo,Robledo_2D-GCM_with_Butler}. 
However, in order to illustrate the robustness of the 2D-GCM 
predictions  with respect to the underlying Gogny-EDF, we will also 
discuss results obtained with the D1S, D1M$^{*}$ and D1M$^{**}$ 
parametrizations for a selected set of nuclei. The parametrization D1S 
has been thoroughly  tested all over the nuclear chart both at the 
mean-field level and beyond (see, for example, 
Ref.~\cite{Review_RoToRa_2019} and references therein). On the other 
hand, D1M$^{*}$ and D1M$^{**}$ are newly proposed re-parametrizations 
of D1M with the goal of improving the slope of the symmetry energy 
while preserving as much as possible other properties  of D1M. Details 
of their fitting protocol can be found in 
Refs.~\cite{gogny-D1MSTAR,gogny-D1MSTARSTAR}.

%

\begin{figure*}
\includegraphics[width=0.95\textwidth]{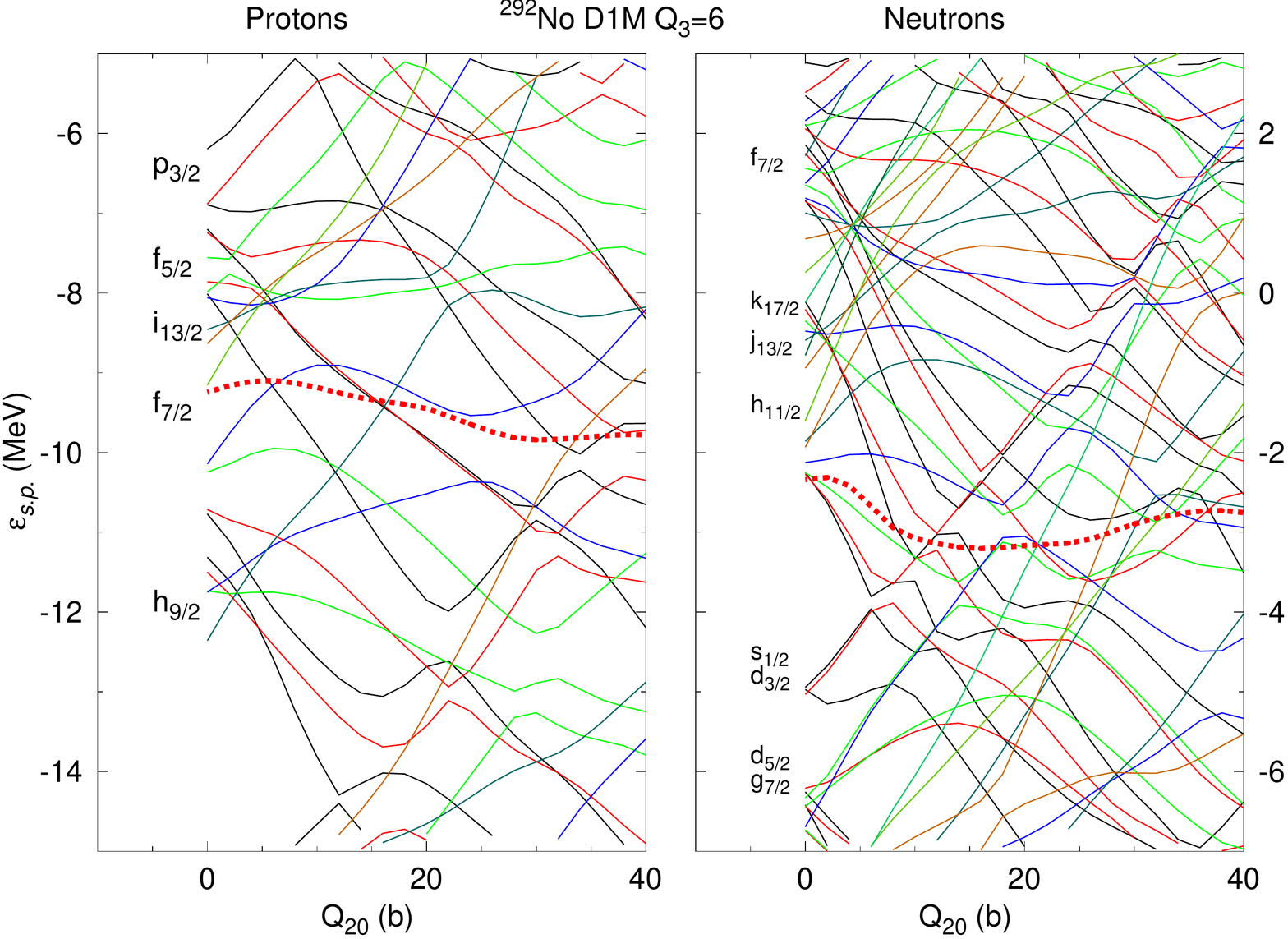}
\caption{(Color online) Same as Fig \ref{MF-spe-RS} but for reflection asymmetric shapes
corresponding to $Q_{30}=6 \textrm{b}^{{3/2}}$. In this case, the single particle
orbitals do not have a definite parity and therefore all of them are plotted with
full lines. 
}
\label{MF-spe-NRS} 
\end{figure*}

The paper is organized as follows. The three levels of approximation 
employed in this study are briefly outlined in 
Secs.~\ref{MF-Theory-used} and \ref{BYMF}. In order to facilitate the 
discussion, the results obtained with the corresponding approach will 
be discussed in each section. The HFB results will be discussed in 
Sec.~\ref{MF-Theory-used} while beyond-mean-field correlations are 
considered in Sec.~\ref{BYMF}. First,  parity-projected potential 
energy surfaces (PPPESs) are computed via parity projection of the 
intrinsic HFB states in Sec.~\ref{PP-Theory-used}. This level of 
approximation is useful to disentangle the relative contribution of 
parity projection to the total correlation. Second, both parity 
projection and fluctuations in the collective coordinates are 
considered via 2D-GCM calculations in Sec.~\ref{2DGCM-Theory-used}. 
Special attention is paid in Sec.~\ref{2DGCM-Theory-used} to the 
systematic of 1$^{-}$ energy splittings, correlation energies as well 
as  $B(E1)$ and $B(E3)$ transition probabilities in the considered 
nuclei. Furthermore, in this section, we will  discuss the robustness 
of the 2D-GCM predictions with respect to the considered Gogny-EDF. 
Finally, Sec.~\ref{conclusions} is devoted to the concluding remarks.

%

\section{Results}

The aim of this work is to study the quadrupole-octupole coupling in 
the neutron-rich nuclei $^{278-294}$U, $^{280-296}$Pu, $^{282-298}$Cm, 
$^{284-300}$Cf, $^{286-302}$Fm, $^{288-304}$No, $^{290-306}$Rf, 
$^{292-308}$Sg, $^{294-310}$Hs and $^{296-312}$Ds. Three levels of 
approximation have been employed: The HFB \cite{rs} scheme with 
constrains on the (axially symmetric) quadrupole  and octupole 
operators, parity projection of the intrinsic state and the symmetry-conserving 2D-GCM. In 
what follows, we briefly outline those approaches, which were used in 
the past in different regions of the nuclear chart 
\cite{Rayner_Q2Q3_GCM_2012,Robledo_2D-GCM_with_Butler,JPG_2020_UPuCmCf_RayRo}. 
The different results obtained at each level of approximation will be 
presented and discussed.

%

\subsection{Mean-field}
\label{MF-Theory-used}

For each of the studied nuclei, we first build the MFPES, i.e., the mean-field energy
$E_{HFB}({\bf{Q}})$ as a function of the $K=0$ multipole deformation moments 
${\bf{Q}} =(Q_{20},Q_{30})$. To this
end, the HFB equation with constrains on the 
axially symmetric quadrupole  
\begin{equation}
\hat{Q}_{20} = z^{2} - \frac{1}{2} \Big(x^{2} + y^{2} \Big)
\end{equation}
and octupole operator
\begin{equation}
\hat{Q}_{30} = z^{3} - \frac{3}{2} \Big(x^{2} + y^{2} \Big)z
\end{equation} 
is solved using an approximate second-order gradient method 
\cite{rob11b} that guarantees a fast and reliable convergence of the selfconsistent HFB procedure. The 
quadrupole $Q_{20}$ and octupole $Q_{30}$ deformation parameters are 
defined via the mean values of the operators $\hat{Q}_{20}$ and 
$\hat{Q}_{30}$ in the corresponding  HFB  states. From the deformations 
$Q_{20}$ and $Q_{30}$, one can easily compute \cite{egi92} the 
deformations parameters $\beta_{2}$ and $\beta_{3}$ as 
\begin{equation}
\beta_{l} = \frac{\sqrt{4 \pi (2l+1)}}{3 R_{0}^{l} A} Q_{l0}
\end{equation}
with $R_{0} = 1.2 A^{1/3}$ and $A$ the mass number. For example, for $A 
= 294$ a quadruple deformation $Q_{20}= 10 \textrm{b}$ is equivalent to 
$\beta_{2} = 0.141$ and an octupole deformation $Q_{30}= 1 \textrm{b}^{3/2}$ is 
equivalent to $\beta_{3} = 0.021$.

%

\begin{figure*}
\includegraphics[angle=-90,width=0.95\textwidth]{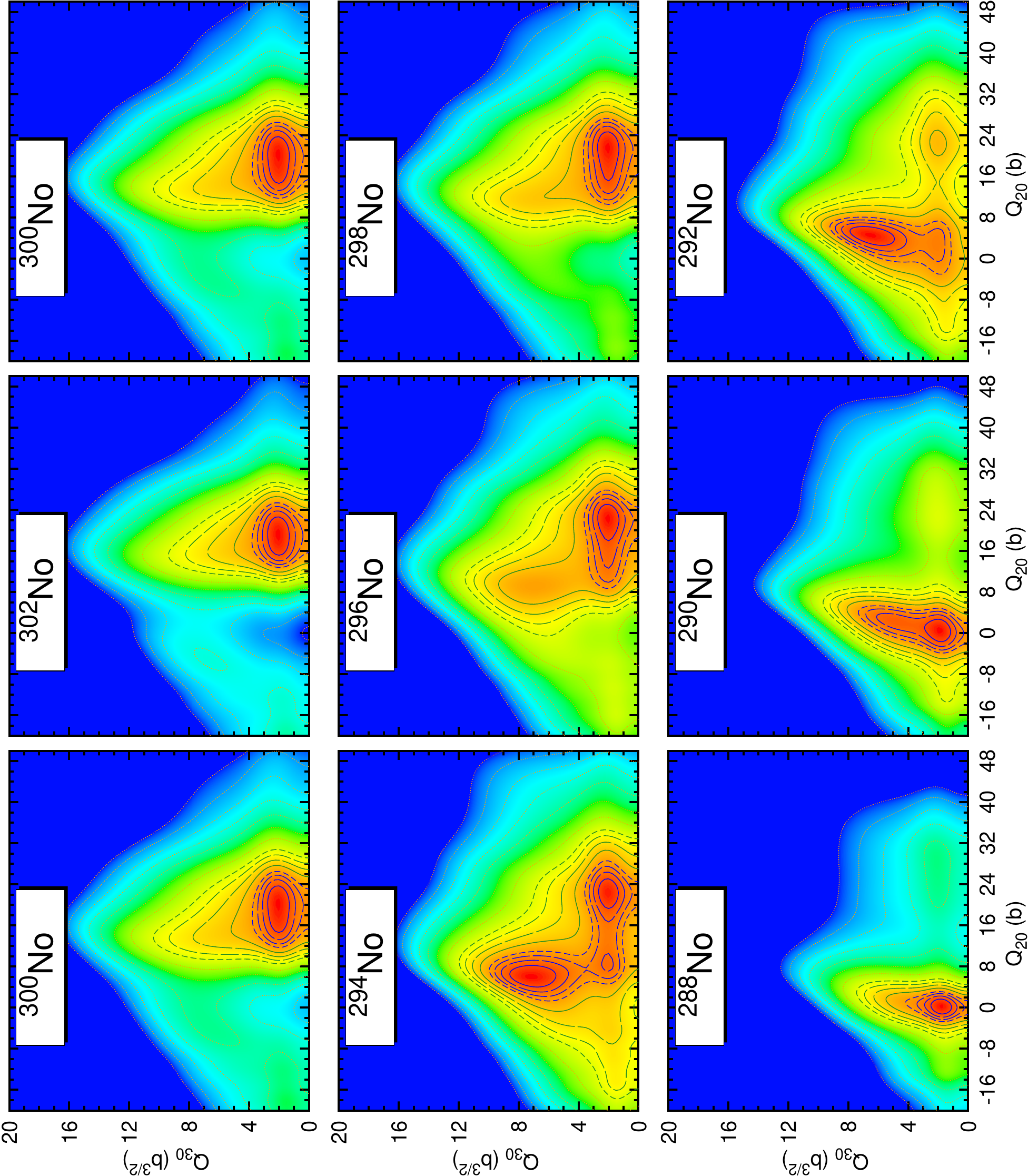}
\caption{(Color online) Positive parity ($\pi = +1$) PPPESs
computed with the Gogny-D1M EDF for the isotopes $^{288-304}$No. See, caption of 
Fig.~\ref{Q2Q3_MF_No} for the contour-line patterns and color scale.
}
\label{Q2Q3_PPLUS_No} 
\end{figure*}

In order to alleviate the already substantial computational effort, 
both axial and time-reversal symmetries have been kept as 
self-consistent symmetries. Aside from the  usual mean-field constrains 
on the proton and neutron numbers, the center of mass is fixed at the 
origin to avoid spurious effects associated with its motion 
\cite{JPG_2012_RoRay,Rayner_Q2Q3_GCM_2012,Robledo_2D-GCM_with_Butler,JPG_2020_UPuCmCf_RayRo}. 
To grant convergence for the studied physical quantities the HFB 
quasiparticle operators $(\hat{\alpha}_{k}^{\dagger},\hat{\alpha}_{k})$ 
\cite{rs} have been expanded in a (deformed)  axially symmetric 
harmonic oscillator (HO) basis $(\hat{c}_{l}^{\dagger},\hat{c}_{l})$ 
containing 17 major shells. 

The MFPESs obtained for the  isotopes $^{288-304}$No  are shown in 
Fig.~\ref{Q2Q3_MF_No}  as illustrative examples. In our calculations, 
the  $Q_{20}$-grid $-20 \textrm{b} \le Q_{20} \le 50 \textrm{b}$ (with 
a step $\delta Q_{20} = 1 \textrm{b} $) and the $Q_{30}$-grid  $0 
\textrm{b}^{3/2} \le Q_{30} \le 20 \textrm{b}^{3/2}$ (with a step 
$\delta Q_{30} = 0.5  \textrm{b}^{3/2}$) have been employed. Along the 
$Q_{20}$-direction there is a shape/phase  transition from a spherical  
ground state in $^{288}$No to well quadrupole-deformed ground states in 
heavier isotopes. For  $^{290-294}$No, the MFPESs exhibit a 
transitional behavior along the $Q_{20}$-direction. Similar results are 
obtained for other isotopic chains. As can be seen from panels 
(a1)-(a5) and (d1)-(d5) of Fig.~\ref{MF-2}, the ground state  
quadrupole deformations $Q_{20,GS}$ are within the range $0 \textrm{b} 
\le Q_{20,GS} \le 30 \textrm{b}$.

The MFPESs show octupole deformed minima in some No isotopes with the 
minima occurring always at small quadrupole deformations. A typical 
example is $^{290}$No which is octupole deformed and almost spherical. 
When the isotopes of No acquire a larger quadrupole deformation the 
octupole deformed minimum vanishes and the ground state becomes 
reflection symmetric with a rather soft MFPES along the 
$Q_{30}$-direction. This pattern repeats in all the other isotopes 
considered in this work as can be deduced from Fig.~\ref{MF-2}. We 
observe there that octupole-deformed HFB ground states are found in 
$^{284-290}$U, $^{284-290}$Pu, $^{286-292}$Cm, $^{286-292}$Cf, 
$^{288-294}$Fm, $^{288-294}$No, $^{292,294}$Rf, $^{294,296}$Sg, 
$^{296}$Hs and $^{298}$Ds with $ 1 \textrm{b}^{3/2} \le Q_{30,GS} \le 7 
\textrm{b}^{3/2}$ [see, panels (b1)-(b5) and (e1)-(e5) of 
Fig.~\ref{MF-2}]. These results indicate that, as in previous studies 
\cite{Agbemava-Q3-2016,Agbemava-Q3-2017,Erler2012,Recent-Survey-Q3,Nix-1995,Jachi-2020}, 
an island of octupole-deformed neutron-rich actinides and low-Z 
superheavy nuclei is found in our HFB calculations based on the  the 
Gogny-D1M EDF. Similar results, not shown here, have  also been 
obtained with the D1S, D1M$^{*}$ and D1M$^{**}$ parametrizations. 
 
%
\begin{figure*}
\includegraphics[angle=-90,width=0.95\textwidth]{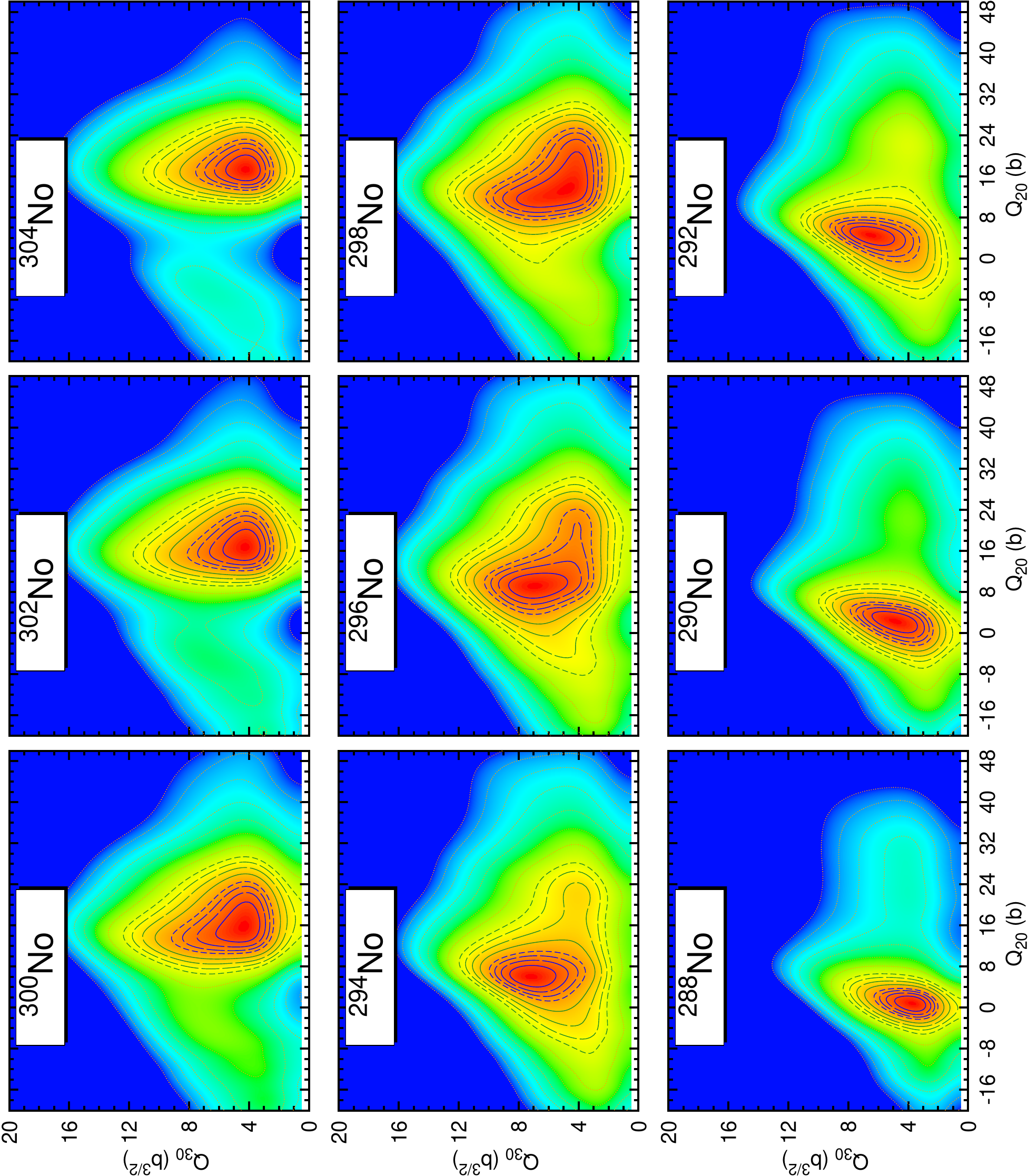}
\caption{(Color online) Negative parity ($\pi = -1$) PPPESs
computed with the Gogny-D1M EDF for the isotopes $^{288-304}$No. See, caption of 
Fig.~\ref{Q2Q3_MF_No} for the contour-line patterns and color scale.
}
\label{Q2Q3_PMINUS_No} 
\end{figure*}

The HFB energy gained by breaking reflection symmetry in the ground 
state, defined as 
\begin{equation}
\label{MFCorrEner}
\Delta E_{CORR, HFB} = E_{HFB, Q_{30}=0} - E_{HFB,GS}
\end{equation}
is plotted in panels (c1)-(c5) and (f1)-(f5) 
of Fig.~\ref{MF-2}.
The largest values of  $\Delta E_{CORR, HFB}$ correspond to 
N =194 (U), N=192 (Pu, Cm, Cf and Fm), N = 190 (No and Rf)
and N = 188 (Sg, Hs and Ds). The maximum value of 1.8 MeV corresponds to
$^{290}$Cf. The relatively small  $\Delta E_{CORR, HFB}$
energies reflect the softness along the $Q_{30}$-direction in the 
MFPESs of nuclei with an octupole-deformed HFB ground state.

For some No isotopes, the MFPESs (see, Fig.~\ref{Q2Q3_MF_No}) exhibit a 
pronounced competition, i.e., shape coexistence, between 
reflection-symmetric and reflection-asymmetric minima. For example, in 
the case of  $^{296}$No the energy difference between the global 
reflection-symmetric $(Q_{20,GS},Q_{30,GS})=(22 \textrm{b}, 0)$ and local 
reflection-asymmetric $(Q_{20},Q_{30})=(10 \textrm{b}, 7 \textrm{b}^{3/2})$ minima 
amounts to just 210 keV. Such a shape coexistence is also observed in other 
isotopic chains.

Before concluding this section, we turn our attention to 
single-particle properties. As it is well known, atomic nuclei "avoid" 
regions with high single-particle level densities (Jahn-Teller effect) 
and therefore the plots of single-particle energies (SPEs) as a 
function of quadrupole or octupole moment help us to identify regions 
where energy gaps (i.e. low level density regions) favor the appearance 
of deformed minima. For this purpose we have chosen to plot the 
eigenvalues of the Routhian $h = t + \Gamma - \lambda_{Q_{20}} Q_{20} - 
\lambda_{Q_{30}} Q_{30}$, where $t$ is the kinetic energy and $\Gamma$ 
the Hartree-Fock field. The term $\lambda_{Q_{20}} Q_{20} + 
\lambda_{Q_{30}} Q_{30}$ contains the Lagrange multipliers used to 
enforce the corresponding quadrupole and octupole constrains.  The 
single-particle energies obtained in the $^{292}$No case are plotted 
in Figs. \ref{MF-spe-RS} and \ref{MF-spe-NRS}, for protons and 
neutrons separately, as  functions of the 
quadrupole moment. The plot of 
Fig.~\ref{MF-spe-RS} corresponds to zero octupole 
deformation, therefore the parity of each single-particle orbital is identified with 
the use of full (positive parity) and dashed (negative parity) lines. 
On the other hand, Fig.~\ref{MF-spe-NRS} corresponds to the same kind of 
plot but in this case we have taken an octupole deformation 
$Q_{30}=6\textrm{b}^{3/2}$ that roughly corresponds with the position of the 
octupole deformed minima. In the later, parity is not a good quantum 
number. Finally, the corresponding Fermi levels are plotted with a 
thick dotted red line. The first thing we notice in Fig.~\ref{MF-spe-RS} is 
the presence of $\Delta j=\Delta l=3$ orbitals 
around the Fermi level both for protons ($i_{13/2}-f_{7/2}$) and 
neutrons ($k_{17/2}-h_{11/2}$). The presence of these opposite parity $\Delta J=\Delta l=3$  pairs of 
orbitals is a natural requirement for the existence 
of octupole-deformed minima. For  protons and neutrons there are gaps in the 
spectra at the spherical configuration relatively close to the Fermi 
level: those gaps are the precursors  of the  near spherical 
octupole-deformed minima observed for neutron numbers N=186-192. The 
deformed minima observed for larger N values at $Q_{20} \approx$ 18 b 
are due to gaps opening up at that deformation. In  Fig.~\ref{MF-spe-NRS} 
we depict the same type of plot but for 
$Q_{30}=6\textrm{b}^{3/2}$. At $Q_{20}=0$ we have included the labels of 
the spherical orbitals at the same place where they are located in
the $Q_{30}=0$ plot to show the strong impact of parity mixing. We 
observe how large shell gaps open up at $Q_{20}=0$ as a consequence of 
parity mixing that are responsible for the near spherical 
octupole-deformed minima obtained for  N=186-192. On the other hand, for 
$Q_{20} \approx$  18 b there are no clear gaps in the spectrum in 
agreement with the fact that there are no octupole-deformed minima for 
that value of the quadrupole moment.

%

\subsection{Beyond-mean-field correlations}
\label{BYMF}

As discussed in the previous section, the softness of the MFPESs along the 
octupole direction as well as the existence in some cases of coexisting 
minima point towards the key role of dynamical beyond-mean-field 
correlations, i.e., symmetry restoration and/or quadrupole-octupole 
configuration mixing in the properties of the ground state and 
collective negative parity states in the studied nuclei. Since the 
octupole is the softest mode, the spatial reflection symmetry is the 
most important invariance to be restored. It would be desirable to restore also 
both the rotational and particle number symmetries. However, such a 
gigantic task is out of the scope of an exhaustive survey like the one 
discussed in this paper for several technical reasons (for example, the 
large number of HO shells used and the number of degrees of freedom 
required in the GCM ansatz).
 
%

\begin{figure}
\includegraphics[width=0.45\textwidth]{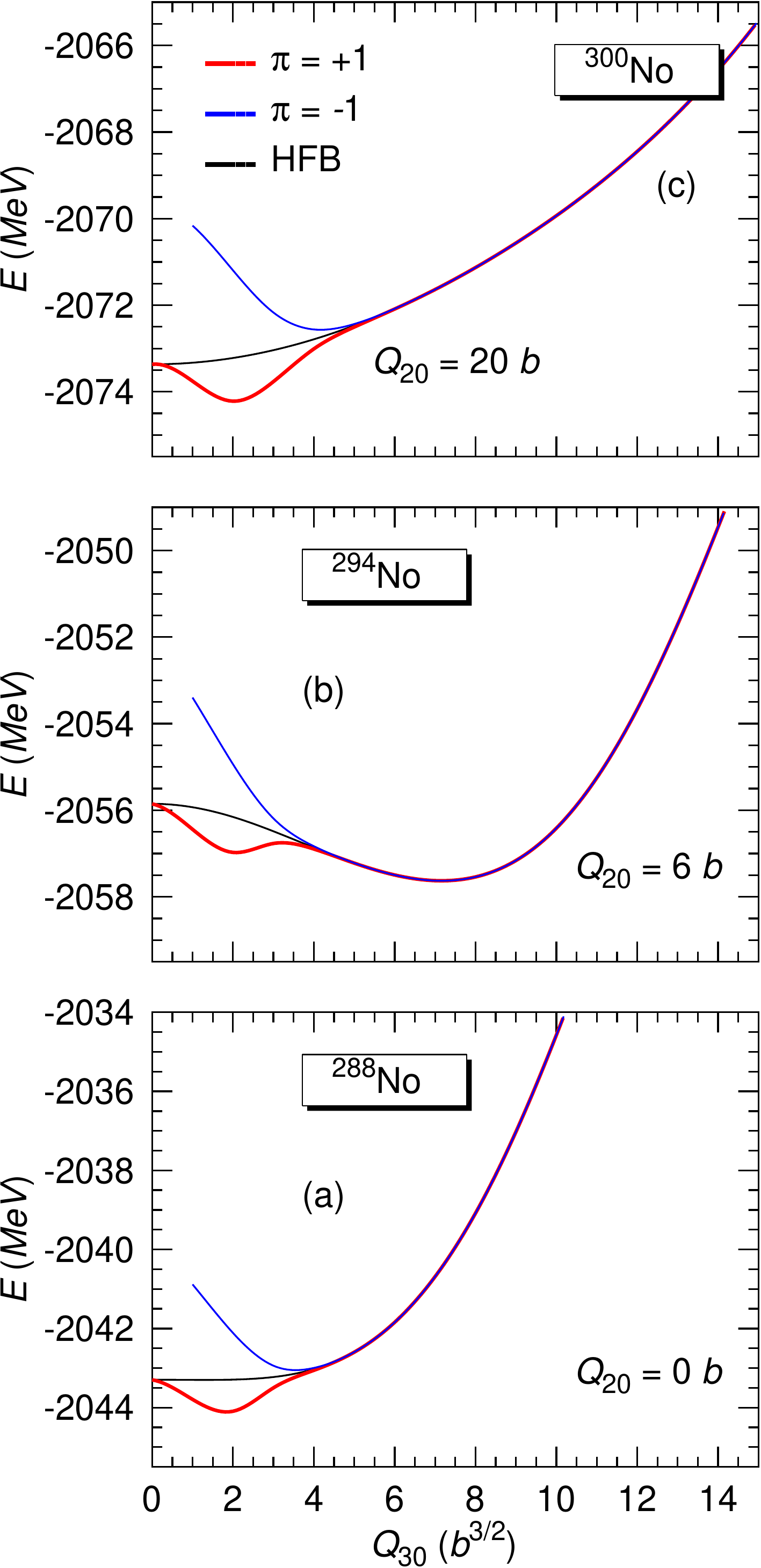}
\caption{(Color online) The $\pi = +1$ (red) and $\pi = -1$ (blue) 
parity-projected energies are depicted as functions of the octupole 
moment $Q_{30}$ for fixed values of the quadrupole moment $Q_{20}$ in 
the nuclei $^{288}$No, $^{294}$No  and $^{300}$No. The corresponding 
HFB energies are also included in the plots. Results have been obtained 
with the Gogny-D1M EDF.
}
\label{examples_No} 
\end{figure}

%

\subsubsection{Parity symmetry restoration}
\label{PP-Theory-used}

In order to restore the spatial reflection symmetry broken by the 
HFB states $| \Phi ({\bf{Q}})\rangle$ with
non-zero octupole deformation, we resort to 
parity projection, i.e., we build 
the parity-conserving 
states  
$| \Phi^{\pi} ({\bf{Q}})\rangle = \hat{{\cal{P}}}^{\pi} | \Phi ({\bf{Q}})\rangle$
by applying on the intrinsic state the parity projector 
\begin{equation}
\hat{{\cal{P}}}^{\pi} = \frac{1}{2} \left(1 + \pi \hat{\Pi} \right),
\end{equation}
where $\pi=\pm 1$ is the desired parity quantum number. The energies 
$E_{\pi} ({\bf Q})$, associated with the states $| \Phi^{\pi} 
({\bf{Q}})\rangle$, define the  PPPESs. They read
\begin{align} \label{PROJEDF}
E_{\pi} ({\bf Q}) &=
\frac{
\langle {\Phi} ({\bf Q}) | \hat{H} [\rho(\vec{r})] | {\Phi} ({\bf Q}) \rangle
}
{2
\langle {\Phi} ({\bf Q}) | \hat{{\cal{P}}}^{\pi} | {\Phi} ({\bf Q}) \rangle
}
\nonumber\\
&+ \pi
\frac{\langle {\Phi} ({\bf Q}) |
\hat{H} [\theta(\vec{r})]  \hat{\Pi} | {\Phi} ({\bf Q}) \rangle
}
{
2
\langle {\Phi} ({\bf Q}) | \hat{{\cal{P}}}^{\pi} | {\Phi} ({\bf Q}) \rangle
}
\nonumber\\
&- \lambda_{Z} 
\Big(
\frac{
\langle {\Phi} ({\bf Q}) | \hat{Z} \hat{{\cal{P}}}^{\pi} | {\Phi} ({\bf Q}) \rangle
}
{\langle {\Phi} ({\bf Q}) | \hat{{\cal{P}}}^{\pi} | {\Phi} ({\bf Q}) \rangle}
-Z_{0}
\Big)
\nonumber\\
&- \lambda_{N} 
\Big(
\frac{
\langle {\Phi} ({\bf Q}) | \hat{N} \hat{{\cal{P}}}^{\pi} | {\Phi} ({\bf Q}) \rangle
}
{\langle {\Phi} ({\bf Q}) | \hat{{\cal{P}}}^{\pi} | {\Phi} ({\bf Q}) \rangle}
-N_{0}
\Big).
\end{align}
Because the Gogny force used is a density dependent one we need a prescription
for the density dependent contribution to the energy overlaps. As in 
previous studies \cite{Rayner_Q2Q3_GCM_2012,Robledo_2D-GCM_with_Butler,JPG_2020_UPuCmCf_RayRo}
we use the density 
\begin{equation}
\rho(\vec{r}) = 
\frac{
\langle {\Phi} ({\bf Q}) | \hat{\rho}({\vec{r})} | {\Phi} ({\bf Q}) \rangle
}
{
\langle {\Phi} ({\bf Q}) | {\Phi} ({\bf Q}) \rangle
}
\end{equation}
to compute $\langle {\Phi} ({\bf Q}) | \hat{H} [\rho(\vec{r})] | {\Phi} ({\bf Q}) \rangle$
and the density 
\begin{equation}
\theta(\vec{r}) = 
\frac{
\langle {\Phi} ({\bf Q}) | \hat{\rho}({\vec{r})} \hat{\Pi} | {\Phi} ({\bf Q}) \rangle
}
{
\langle {\Phi} ({\bf Q}) | \hat{\Pi} | {\Phi} ({\bf Q}) \rangle
}
\end{equation}
in the evaluation of $\langle {\Phi} ({\bf Q}) | \hat{H} 
[\theta(\vec{r})]  \hat{\Pi} | {\Phi} ({\bf Q}) \rangle$. In this way 
we avoid the pathologies found in the restoration of spatial symmetries 
\cite{rod02,egi04,robledo_presciption-1,robledo_presciption-2}. As the 
parity-projected  proton and neutron numbers, usually differ from the 
nucleus' proton $Z_{0}$ and neutron $N_{0}$ numbers, we have introduced 
first-order corrections in Eq. (\ref{PROJEDF}), with $\lambda_{Z}$ and 
$\lambda_{N}$ being chemical potentials for protons and neutrons, 
respectively 
\cite{har82,bon90,Rayner_Q2Q3_GCM_2012,Robledo_2D-GCM_with_Butler,JPG_2020_UPuCmCf_RayRo}
. 

%

\begin{figure*}
\includegraphics[angle=-90,width=0.95\textwidth]{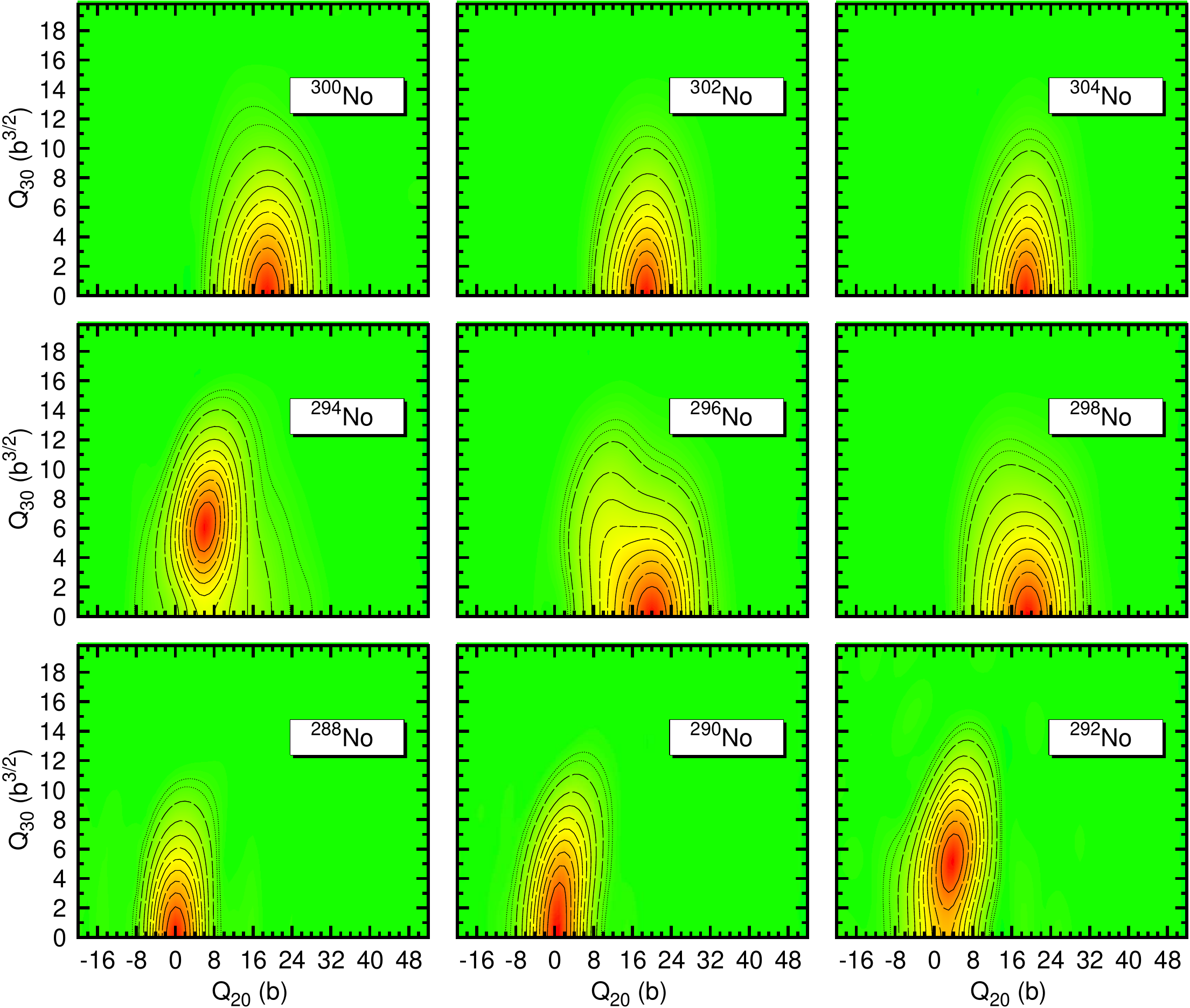}
\caption{Collective wave functions Eq. (\ref{cll-wfs-HW}) for 
the ground states of the nuclei $^{288-304}$No. The contour lines (a 
succession of solid, long dashed and short dashed lines) start at $90 
\%$ of the maximum value up $10 \%$ of it. The two dotted-line contours 
correspond to the tail of the amplitude ($5 \%$ and $1 \%$ of the 
maximum value). The color scale ranges from red (maximum value) to green
(zero). Results have been obtained with the Gogny-D1M EDF. For 
more details, see the main text.
}
\label{COLLWS_POS_PARITY_No} 
\end{figure*}

The $\pi =+1$ and $\pi =-1$ PPPESs obtained for the  isotopes 
$^{288-304}$No  are depicted in Figs.~\ref{Q2Q3_PPLUS_No} and 
\ref{Q2Q3_PMINUS_No} as illustrative examples. Since $\hat{\Pi} | \Phi 
(Q_{20},Q_{30}=0)\rangle = | \Phi (Q_{20},Q_{30}=0)\rangle $, the 
projection onto positive parity is unnecessary for those states. On the 
other hand, in the case of negative parity, the evaluation of the 
projected energy along the $Q_{30}=0$ axis requires to resolve 0/0 
indeterminacy \cite{egi91} and therefore it is subject to numerical 
inaccuracies \cite{Rayner_Q2Q3_GCM_2012,JPG_2020_UPuCmCf_RayRo}. 
However, the negative parity projected energy $E_{\pi=-1} ({\bf Q})$ 
Eq. (\ref{PROJEDF}) increases rapidly when approaching $Q_{30}=0$ (see, 
Fig.~\ref{examples_No}) 
\cite{Rayner_Q2Q3_GCM_2012,JPG_2020_UPuCmCf_RayRo} and  its limiting 
value \cite{egi91} is high enough as not to play a significant role in 
the discussion of the $\pi =-1$ PPPESs. We have then omitted this 
quantity along the $Q_{30}=0$ axis in Fig.~\ref{Q2Q3_PMINUS_No}. It is 
worth to notice that the quadrupole moments corresponding to the 
absolute minima of the $\pi =+1$ and $\pi =-1$ PPPESs are close to the 
HFB values.

As can be seen from Figs.~\ref{Q2Q3_MF_No}, \ref{Q2Q3_PPLUS_No} and 
\ref{examples_No}, not only the MFPESs but also the $\pi =+1$ PPPESs 
are rather soft along the $Q_{30}$-direction. In the case of nuclei 
with small and/or zero HFB ground state octupole deformations, such as 
$^{288}$No and $^{300}$No, the $\pi =+1$ PPPESs only display an 
absolute minimum around $Q_{30} = 2.0 \textrm{b}^{3/2}$. This is illustrated in 
panels (a) and (c) of  Fig.~\ref{examples_No} where the $\pi =+1$ 
parity-projected energies obtained for $^{288}$No and  $^{300}$No are 
plotted, as functions of $Q_{30}$,  for a fixed value of the quadrupole 
moment corresponding  to the absolute minimum of the PES. However, the 
topography along the $Q_{30}$-direction is more complex for nuclei with 
larger HFB  octupole deformations as  the $\pi =+1$ PPPESs exhibit a 
pronounced competition between two minima. In the case of $^{294}$No, 
for example, the energy difference between the local $Q_{30} = 2.0 
\textrm{b}^{3/2}$ and global $Q_{30} = 7.0 \textrm{b}^{3/2}$ minima [see, panel (b) of  
Fig.~\ref{examples_No}] amounts to 652 keV. Note also from panel (b) of  
Fig.~\ref{examples_No}, the energy degeneracy of the absolute HFB and  
$\pi =+1$ minima in this case. Furthermore, as can be seen from 
Fig.~\ref{Q2Q3_PPLUS_No}, for $^{294}$No the shape coexistence extends 
to a third minimum, located at $(Q_{20},Q_{30}) = (22\textrm{b}, 2\textrm{b}^{3/2})$, 
which is only 14 keV above the absolute one.

%
\begin{figure*}
\includegraphics[angle=-90,width=0.95\textwidth]{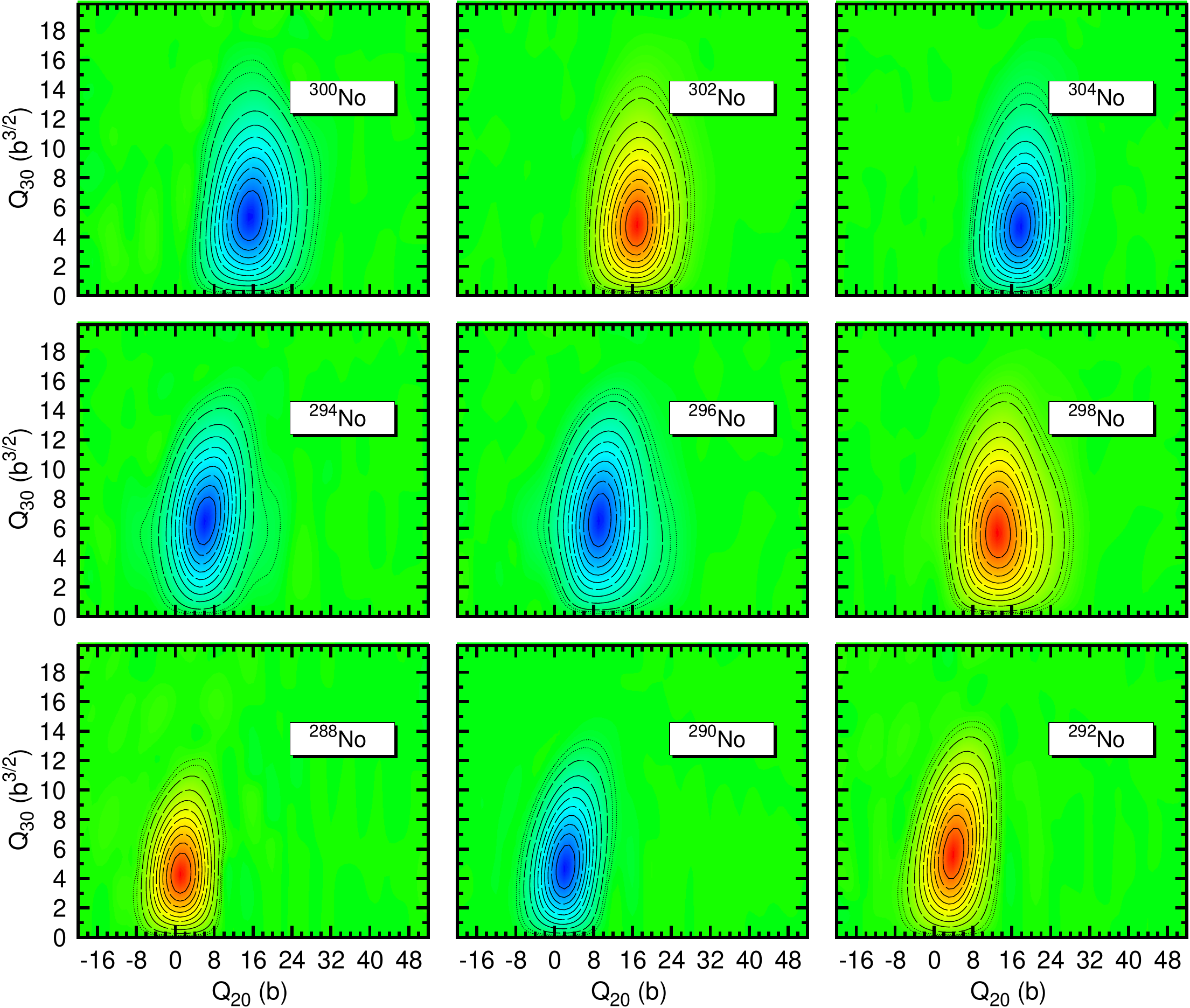}
\caption{(Color online) Collective wave functions Eq. (\ref{cll-wfs-HW}) for 
the lowest negative-parity states of the nuclei $^{288-304}$No. See, 
caption of Fig.~\ref{COLLWS_POS_PARITY_No} for contour-line patterns. 
Results have been obtained with the Gogny-D1M EDF. For more details, 
see the main text. 
}
\label{COLLWS_NEG_PARITY_No} 
\end{figure*}

The $\pi =-1$ PPPESs, depicted in \ref{Q2Q3_PMINUS_No}, exhibit in all 
the cases a well developed absolute minimum. In the case of nuclei such 
as $^{288}$No and $^{300}$No, the absolute $\pi =-1$ minimum  
corresponds to a larger octupole deformation than the $\pi =+1$  one 
[see, panels (a) and (c) of Fig.~\ref{examples_No}]. On the other hand, 
for $^{294}$No, the (degenerate) $\pi =-1$  and $\pi =+1$ absolute 
minima  have similar octupole deformations [see, panel (b) of 
Fig.~\ref{examples_No}]. Similar features have been found for other 
isotopic chains. Let us mention, that the complex topography found for 
the PPPESs along the $Q_{30}$-direction in our Gogny-D1M calculations 
has already  been studied, as a function of the strength of the 
two-body interaction, using parity-projection on the  
Lipkin-Meshkov-Glick (LMG) model \cite{LMG-model}.

As a measure of the correlations induced by parity projection, we  
consider the correlation energy
\begin{equation}
\label{PPCorreEner}
\Delta E_{CORR, PP} = E_{HFB, GS} - E_{\pi=+1,GS}.
\end{equation}
defined in terms of the difference between the HFB  $E_{HFB, GS}$ and 
parity-projected $E_{\pi=+1,GS}$  ground state energies. In 
Fig.~\ref{corre_energies}, we show this quantity for the studied 
nuclei. The correlation energy $\Delta E_{CORR, PP}$ is zero or nearly 
zero for  U, Pu, Cm, Cf, Fm and No isotopes with  $ 190 \le N \le 194$ 
as for these nuclei the $\pi =+1$ PPPESs display features, along the 
$Q_{30}$-direction, similar to the ones discussed above for $^{294}$No, 
i.e., the HFB and $\pi =+1$ absolute minima are degenerate or nearly 
degenerate. As  will be discussed in Sec \ref{2DGCM-Theory-used}, the 
comparison between the correlation energies $\Delta E_{CORR, PP}$ and 
the ones obtained within the symmetry-conserving 2D-GCM framework 
reveals the key role played by quantum fluctuations around those 
neutron numbers. 

%

\subsubsection{Generator Coordinate Method}
\label{2DGCM-Theory-used}

The dynamical interplay between the quadrupole and octupole degrees of
freedom is considered via the  2D-GCM ansatz
\begin{equation} \label{GCM-WF}
| {\Psi}_{\sigma}^{\pi} \rangle = \int d{\bf Q} f_{\sigma}^{\pi} ({\bf Q}) | {\Phi} ({\bf Q}) \rangle
\end{equation}
where, both positive and negative 
octupole moments  $Q_{30}$ are included in the integration domain to
assure the parity-conserving nature of the states  $| {\Psi}_{\sigma}^{\pi} \rangle$
\cite{Rayner_Q2Q3_GCM_2012,JPG_2020_UPuCmCf_RayRo,Robledo_2D-GCM_with_Butler}. 
The index $\sigma$ in Eq. (\ref{GCM-WF}) labels the different GCM solutions.

%
\begin{figure*}
\includegraphics[angle=-90,width=0.95\textwidth]{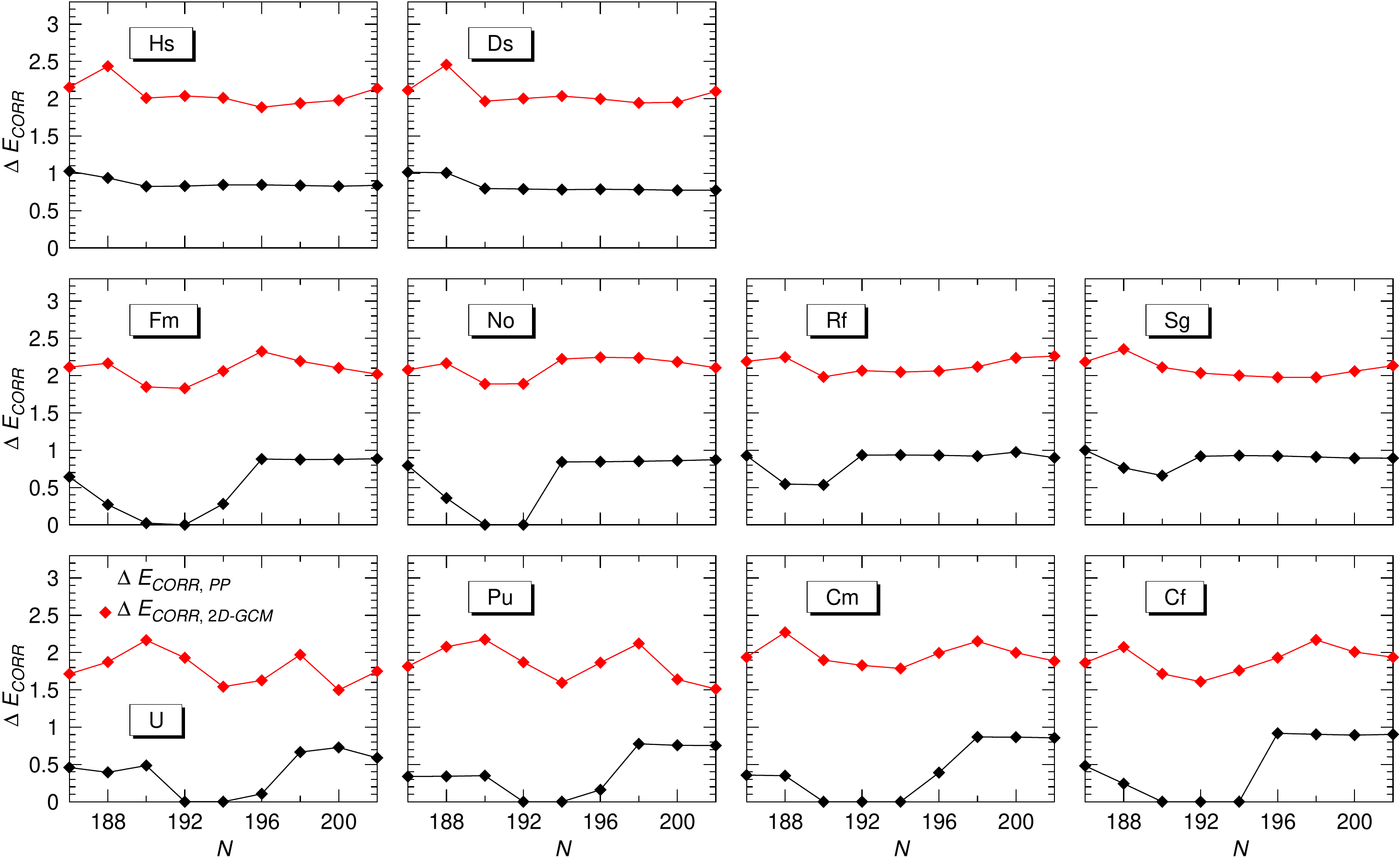}
\caption{(Color online) The correlation energies obtained within the 
2D-GCM framework Eq. (\ref{2DGCMCorreEner}) are plotted as functions of 
the neutron number. The correlation energies $\Delta E_{CORR, PP}$ 
Eq. (\ref{PPCorreEner}) stemming from the restoration of  reflection 
symmetry are also included in the plots. Results have been obtained 
with the Gogny-D1M EDF. For more details, see the main text. 
}
\label{corre_energies} 
\end{figure*}

The amplitudes $f_{\sigma}^{\pi} ({\bf Q})$ are solutions of the 
Griffin-Hill-Wheeler (GHW) equation \cite{rs}
\begin{equation} \label{HW-equation}
\int d{\bf Q}^{'} 
\left(
{\cal{H}}({\bf Q},{\bf Q}^{'}) - E_{\sigma}^{\pi}
{\cal{N}}({\bf Q}, {\bf Q}^{'})
\right)
f_{\sigma}^{\pi} ({\bf Q}^{'}) = 0
\end{equation} 
with the Hamiltonian and norm kernels  given by
\begin{align} \label{GCM-PROJEDF-hnk}
{\cal{H}}({\bf Q}, {\bf Q}^{'}) &=
\langle {\Phi} ({\bf Q}) | 
\hat{H} [\rho^{GCM}(\vec{r}) ] | {\Phi} ({\bf Q}^{'}) \rangle
\nonumber\\
&- \lambda_{Z}  \Big( \langle {\Phi} ({\bf Q}) | \hat{Z} | {\Phi} ({\bf Q}^{'}) \rangle
- Z_{0} {\cal{N}}({\bf Q}, {\bf Q}^{'})\Big)
\nonumber\\
&- \lambda_{N}  \Big( \langle {\Phi} ({\bf Q}) | \hat{N} | {\Phi} ({\bf Q}^{'}) \rangle
- N_{0} {\cal{N}}({\bf Q}, {\bf Q}^{'})\Big)
\nonumber\\
{\cal{N}}({\bf Q}, {\bf Q}^{'}) 
&= \langle {\Phi} ({\bf Q}) |  {\Phi} ({\bf Q}^{'}) \rangle.
\end{align}
As in previous studies \cite{Rayner_Q2Q3_GCM_2012,Robledo_2D-GCM_with_Butler,JPG_2020_UPuCmCf_RayRo}
we use the {\it{mixed}} density prescription
\begin{equation}
\rho^{GCM}(\vec{r})= 
\frac{
\langle {\Phi} ({\bf Q}) | \hat{\rho}({\vec{r})} | {\Phi} ({\bf Q}^{'}) \rangle
}
{
\langle {\Phi} ({\bf Q}) | {\Phi} ({\bf Q}^{'}) \rangle
}
\end{equation} 
in the density dependent term of the Hamiltonian kernel. As in the 
parity projection case Eq. (\ref{PROJEDF}), we use a perturbative 
first-order correction in the Hamiltonian  kernel ${\cal{H}}({\bf Q}, 
{\bf Q}^{'})$ to take into account deviations in both the proton and 
neutron numbers 
\cite{har82,bon90,Rayner_Q2Q3_GCM_2012,Robledo_2D-GCM_with_Butler}.

The HFB basis intrinsic states $| {\Phi} ({\bf Q}) \rangle$ are not
orthonormal. Therefore, the amplitudes $f_{\sigma}^{\pi} ({\bf Q})$ 
cannot be interpreted as probability amplitudes. 
Instead, one considers the so-called collective wave functions 
\begin{equation} \label{cll-wfs-HW} 
G_{\sigma}^{\pi} ({\bf Q}) =   \int d{\bf Q}^{'} {\cal
{N}}^{\frac{1}{2}}({\bf Q}, {\bf Q}^{'})  f_{\sigma}^{\pi}({\bf 
Q}^{'}),
\end{equation}  
written in terms of the operational square root 
of the norm kernel ${\cal
{N}}^{\frac{1}{2}}({\bf Q}, {\bf Q}^{'})$ 
\cite{rs,rod02,Rayner_Q2Q3_GCM_2012,Robledo_2D-GCM_with_Butler,JPG_2020_UPuCmCf_RayRo}. 

The overlaps of one and two body operators between different HFB states
are evaluated with the efficient pfaffian techniques of Refs \cite{rob09,rob11,ber12}. 

%
 
\begin{figure*}
\includegraphics[width=0.95\textwidth]{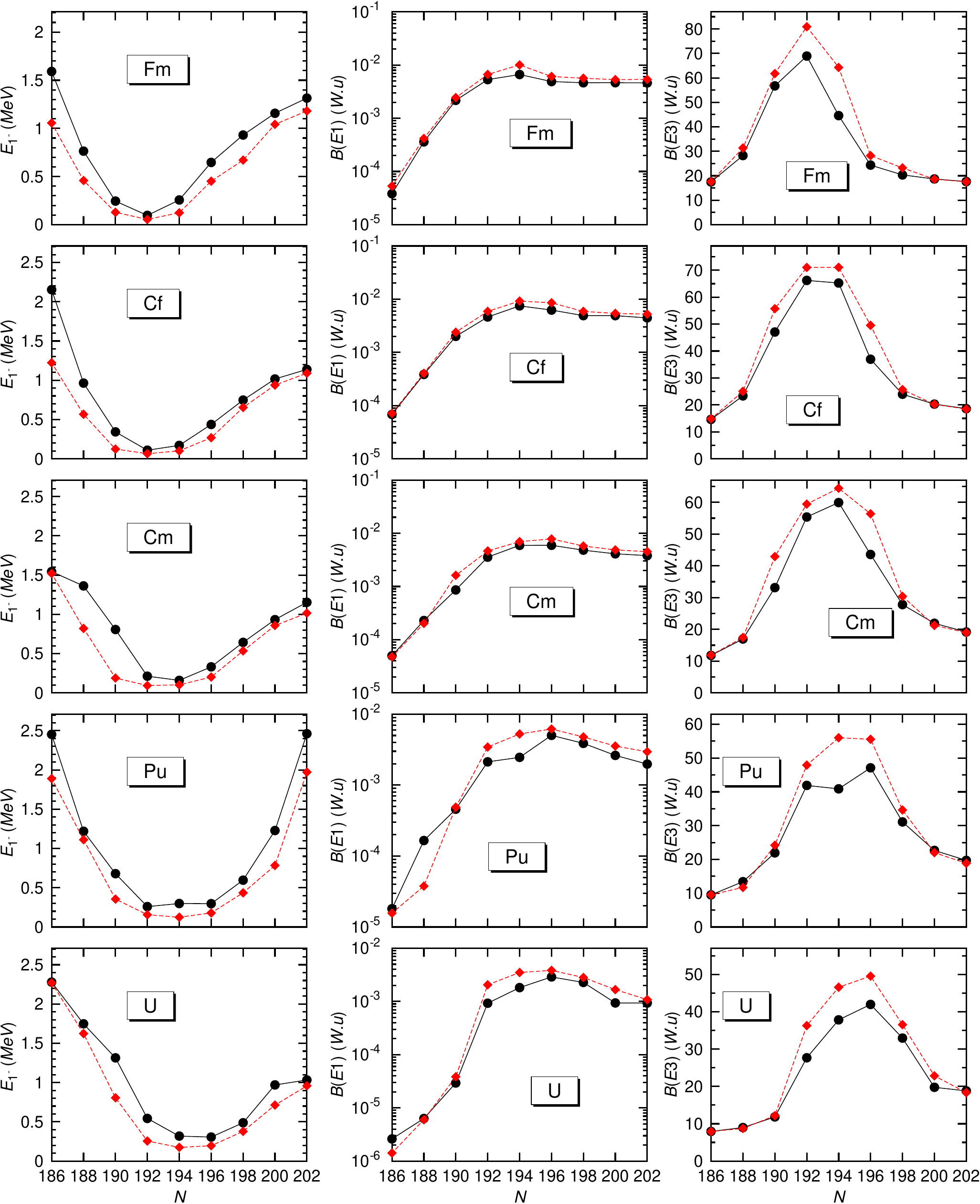}
\caption{(Color online) The 2D-GCM $E_{1^{-}}$ energy splittings (left 
panels) and the  reduced transition probabilities $B(E1)$ (middle 
panels) and $B(E3)$ (right panels) are plotted (in black) as functions 
of the neutron number for the studied U, Pu, Cm, Cf and Fm isotopic 
chains. The $E_{1^{-}}$, $B(E1)$ and $B(E3)$ values obtained in the 
framework of the 1D-GCM, with the octupole moment as single  generating 
coordinate, have also been included (in red) in each of the plots. 
Results have been obtained with the Gogny-D1M EDF. For more details, 
see the main text.
}
\label{E1BE1BE3-GCM-1} 
\end{figure*}

For the reduced transition probabilities 
$B(E1,1^{-} \rightarrow 0^{+} )$ and 
$B(E3,3^{-} \rightarrow 0^{+} )$ the rotational 
formula for K=0 bands has been used 
\begin{equation} \label{ROT_FORMULA}
	B(E\lambda,\lambda^{-} \rightarrow 0^{+} ) = \frac{e^{2}}{4 \pi} 
	\Big{|} \langle \Psi_{\sigma}^{\pi=-1} | \hat{{\cal O}}_{\lambda} 
	|\Psi_{{\sigma}^{'}=1}^{{\pi}^{'}=+1} \rangle \Big{|}^{2}.
\end{equation}
For $B(E1)$ and $B(E3)$ transitions $\sigma$ corresponds to the first 
excited GCM state with negative parity. The electromagnetic transition 
operators $\hat{{\cal O}}_{1}$ and $\hat{{\cal O}}_{3}$ are the dipole 
moment operator and the proton component of the octupole operator, 
respectively \cite{Rayner_Q2Q3_GCM_2012}. The overlap $ \langle 
\Psi_{\sigma}^{\pi} | \hat{O} |\Psi_{\sigma '}^{\pi '} \rangle$ of an 
operator $\hat{O}$ between two different GCM states Eq. (\ref{GCM-WF}) 
can be evaluated according to the expressions given in 
Refs.~\cite{Rayner_Q2Q3_GCM_2012,JPG_2020_UPuCmCf_RayRo}.

The collective wave functions Eq. (\ref{cll-wfs-HW}) 
corresponding to the ground and lowest negative-parity 2D-GCM 
states in $^{288-304}$No
are plotted  
in Figs.~\ref{COLLWS_POS_PARITY_No} 
and \ref{COLLWS_NEG_PARITY_No}, respectively. As can be seen from
Fig.~\ref{COLLWS_POS_PARITY_No},  
the ground state collective amplitude
$G_{\sigma=1}^{\pi=+1} (Q_{20},Q_{30})$ shows the typical Gaussian shape
along both the quadrupole and octupole directions with a maximum located 
at octupole moments  different from 
zero in $^{292,294}$No. The same holds for 
$^{284,286}$U, $^{286-290}$Pu, $^{286-290}$Cm, $^{288-292}$Cf,
$^{290,292}$Fm, $^{294}$Rf and $^{296}$Sg.  For other nuclei
the peaks of the ground state collective amplitudes
are located around $Q_{30} =0$. The spreading of 
$G_{\sigma=1}^{\pi=+1} (Q_{20},Q_{30})$
along the $Q_{30}$-direction is large, indicating the 
octupole-soft character of the 2D-GCM ground 
states obtained for the considered nuclei. 
On the other hand, for the negative parity amplitudes 
$G_{\sigma}^{\pi=-1} (Q_{20},Q_{30})$, depicted in Fig.~\ref{COLLWS_NEG_PARITY_No}, the
shape of the wave function is again Gaussian along the $Q_{20}$-direction
whereas along the $Q_{30}$-direction it shows the characteristic shape of the 
first excited state of the harmonic oscillator (odd under the exchange of
sign in $Q_{30}$) with a zero value at
$Q_{30}=0$ as well as a maximum and a minimum, one at a positive $Q_{30}$ value 
and the other at the corresponding negative value. 
As a consequence, the negative parity wave function maximum or minimum 
always take place at a nonzero octupole moment. This is in agreement with
the position of the minima  of the $\pi=-1$ PPPESs (see, Fig.~\ref{Q2Q3_PMINUS_No}).
 
The 2D-GCM average quadrupole moment is defined as
\begin{equation}
	(\bar{Q}_{20})_{\sigma}^{\pi}= 
	\langle {\Psi}_{\sigma}^{\pi} | \hat{Q}_{20}| {\Psi}_{\sigma}^{\pi} \rangle
\end{equation}	
and the ground state values $(\bar{Q}_{20})_{\sigma=1}^{\pi=+1}$, corroborate 
the mean field 
result, i.e., with increasing neutron number, for each of the studied 
isotopic chains, there is a transition to well quadrupole-deformed 
ground states. Similarly, the average quadrupole moments of the first 
negative parity excited state $(\bar{Q}_{20})_{\sigma}^{\pi=-1}$, 
increase  with increasing neutron number.  

%

\begin{figure*}
\includegraphics[width=0.95\textwidth]{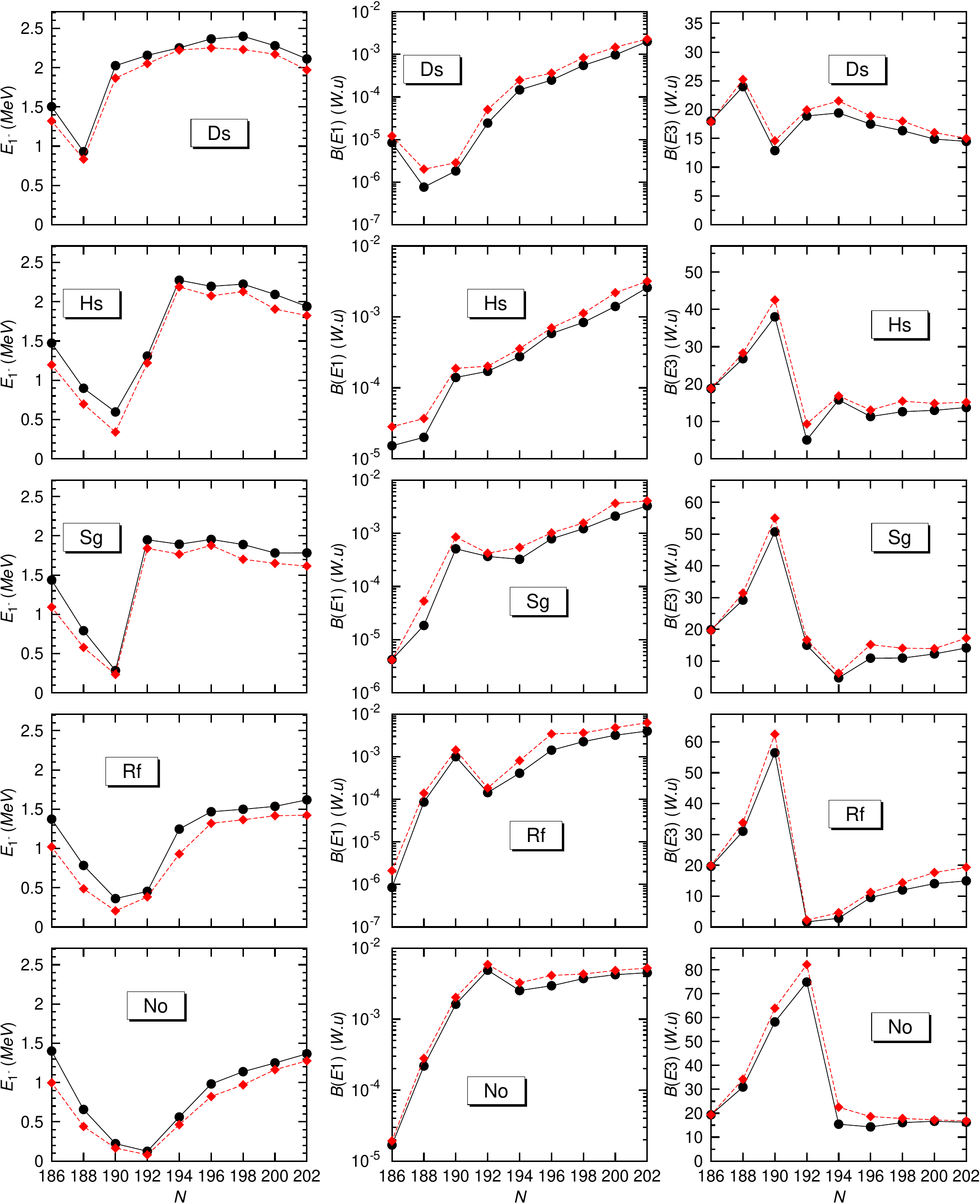}
\caption{(Color online) The same as Fig.~\ref{E1BE1BE3-GCM-1} but for 
the No, Rf, Sg, Hs and Ds isotopic chains.
}
\label{E1BE1BE3-GCM-2} 
\end{figure*} 

We have also computed the average octupole moment
\cite{Rayner_Q2Q3_GCM_2012,JPG_2020_UPuCmCf_RayRo}
\begin{equation}
	(\bar{Q}_{30})_{\sigma}^{\pi}= 4 \int_{\mathcal{D}} d {\bf Q} d {\bf Q'} G_{\sigma}^{\pi \,*} ( {\bf Q} ) 
	 {\cal Q }_{30} ({\bf Q}, {\bf Q}^{'})
G_{\sigma}^{\pi} ({\bf Q}^{'})
\end{equation}
and obtained, for all the studied nuclei, non zero values in the range 
$0.37 b^{3/2} \le (\bar{Q}_{30})_{\sigma=1}^{\pi=+1} \le 5.41 b^{3/2}$. 
At variance with the static  HFB results of Sec.~\ref{MF-Theory-used}, 
once parity-projected quadrupole-octupole configuration mixing effects 
are taken into account via the 2D-GCM ansatz Eq. (\ref{GCM-WF}), the 
ground states of all the studied nuclei are (dynamically) octupole-deformed 
albeit with the largest octupole deformations  
$(\bar{Q}_{30})_{\sigma=1}^{\pi=+1}$ corresponding to U, Pu, Cm, Cf, 
Fm, No, Rf and Sg isotopes with $190 \le N \le 196$. For the  octupole 
moments $(\bar{Q}_{30})_{\sigma}^{\pi=-1}$  we have obtained values in 
the range $2.57 b^{3/2} \le (\bar{Q}_{30})_{\sigma}^{\pi=-1} \le 6.17 
b^{3/2}$ and their largest values   correspond once more to U, Pu, Cm, 
Cf, Fm, No, Rf and Sg isotopes with $190 \le N \le 196$.

The 2D-GCM correlation energy 
\begin{equation}
\label{2DGCMCorreEner}
\Delta E_{CORR, 2D-GCM} = E_{HFB, GS} - E_{\pi=+1,2D-GCM}
\end{equation}
is defined as the difference between the  HFB and 2D-GCM ground-state 
energies. This quantity is plotted  in Fig.~\ref{corre_energies} along 
with the correlation energy $\Delta E_{CORR, PP}$ stemming from 
symmetry restoration alone . The comparison between both correlation 
energies reveals that 2D-GCM zero-point quantum fluctuations 
substantially modify the behavior of $\Delta E_{CORR, PP}$ for U, Pu, 
Cm, Cf, Fm and No isotopes with $190 \le N \le 196$ providing a weaker 
dependence of $\Delta E_{CORR, 2D-GCM}$ with the neutron number. A 
weaker trend is also obtained for Rf and Sg nuclei around N=190. This 
agrees well with previous results for Sm, Gd and actinide nuclei 
\cite{Rayner_Q2Q3_GCM_2012,JPG_2020_UPuCmCf_RayRo}. Moreover, the range 
of values of the correlation energy $1.49 MeV  \le \Delta E_{CORR, 
2D-GCM} \le 2.45 MeV $ is of the same order of magnitude as the rms for 
the binding energy in Gogny-like nuclear mass tables \cite{gogny-d1m} 
and, therefore, those correlation energies should be considered in 
future parametrizations of the Gogny-EDF.

%
\begin{figure*}
\includegraphics[angle=-90,width=0.95\textwidth]{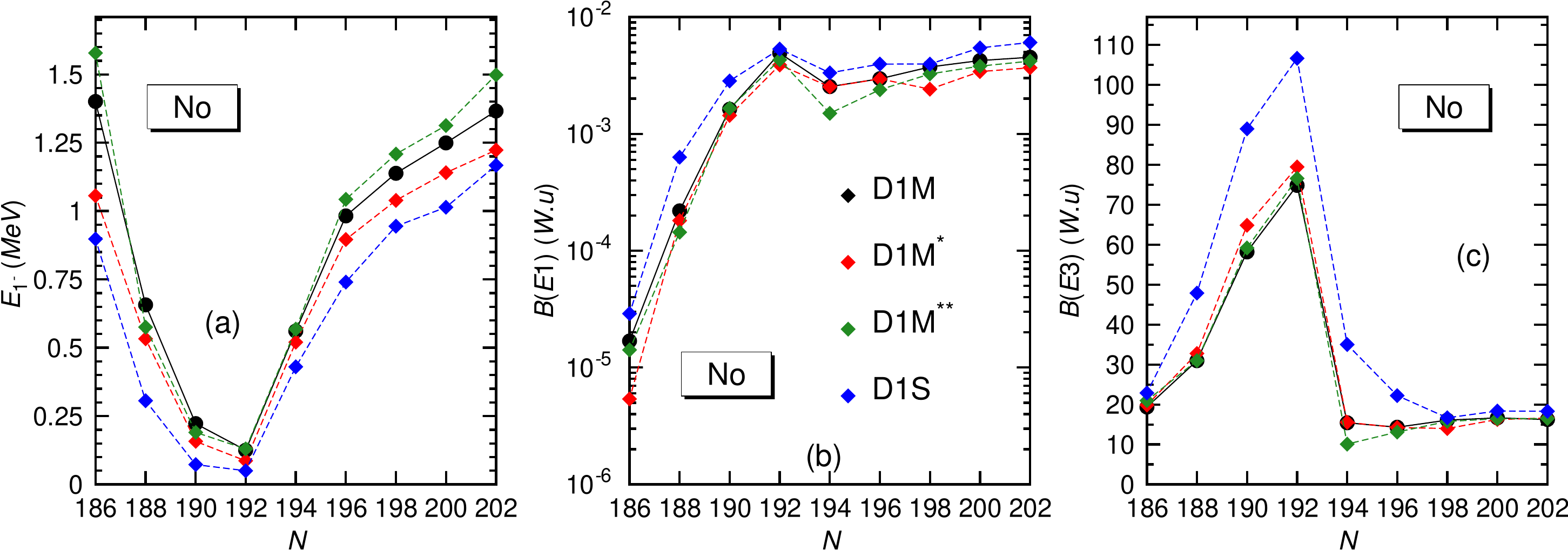}
\caption{(Color online) The 2D-GCM $E_{1^{-}}$ energy splittings [panel 
(a)] and the  reduced transition probabilities $B(E1)$ [panel (b)] and 
$B(E3)$ [panel (c)] are plotted as functions of the neutron number for 
the isotopes $^{288-304}$No. Results have been obtained with the 
parametrizations D1S, D1M, D1M$^{*}$ and D1M$^{**}$ of the Gogny-D1M 
EDF. For more details, see the main text.
}
\label{E1BE1BE3-GCM-sensitivity} 
\end{figure*}

The 2D-GCM   energy difference between the positive-parity ground state 
and the lowest $1^-$ excited state is depicted in the left panels of 
Figs.~\ref{E1BE1BE3-GCM-1}  and \ref{E1BE1BE3-GCM-2} as a function of 
the neutron number. The $1^{-}$ excitation energies are very small $( 
0.10 MeV \le E_{1^{-}} \le 0.36 MeV)$ for U, Pu, Cm, Cf, Fm, No, Rf and 
Sg isotopes with $190 \le N \le 196$ in agreement with their large 
dynamical octupole deformations. Note, that in the case of Hs and Ds 
isotopes, the $E_{1^{-}}$ values obtained for $^{298}$Hs (0.60 MeV) and 
$^{298}$Ds (0.93 MeV) are slightly larger than the previous ones. Other 
nuclei, with less pronounced dynamical octupole deformation effects, 
exhibit larger $E_{1^{-}}$ values pointing towards the octupole 
vibrational character of their  first negative-parity excited state. In 
the same panels, we have also included the $E_{1^{-}}$ energies 
obtained within 1D-GCM calculations with $Q_{30}$ as single generating 
coordinate. It is satisfying to observe that both calculations predict 
very similar trends with neutron number though the 2D-GCM energies are  
larger the the 1D-GCM ones.

The $B(E1)$ transition probabilities are plotted in the middle panels 
of Figs.~\ref{E1BE1BE3-GCM-1}  and \ref{E1BE1BE3-GCM-2}. For 
92 $\le$ Z $\le$ 102, they exhibit a 
steady increase up to N=192 while for larger neutron numbers, the
$B(E1)$ strengths remain almost constant. Exception made of $^{298,300}$Ds, a
steady increase is 
also observed for Z $\ge$ 104 up to N=190. At variance with the results 
obtained for lower-Z chains, exception made of  $^{296}$Rf and $^{298,300}$Sg, the
$B(E1)$ values  also increase for larger neutron numbers, being the effect  
more pronounced in the Hs and Ds isotopic chains. Note, that the behavior 
of the B(E1) strengths with neutron number is not 
correlated with the behavior of the negative parity excitation energies 
and the B(E3) strengths (discussed below). Therefore, it  is not strictly 
correlated with the amount of octupole correlations. This is a consequence 
of the strong dependence of the dipole moment with orbital 
occupancies \cite{egi90} that leads, for instance, to strong 
suppression of the E1 strength in some specific nuclei 
\cite{egi89,egi92} and not in their neighbors. As can be seen
from  the figures, the $B(E1)$ transition probabilities obtained 
within 1D-GCM calculations exhibit the same pattern with 
neutron number as the 2D-GCM ones.

The $B(E3)$ transition probabilities are plotted in the right panels of 
Figs.~\ref{E1BE1BE3-GCM-1}  and \ref{E1BE1BE3-GCM-2}. Contrary to the 
B(E1) case, the magnitude of the B(E3) strength is strongly correlated 
with the excitation energy of the collective negative parity state, i.e.,  
whenever this excitation energy is small the B(E3) strength is large.
Exception made of $^{296}$No and $^{298}$Rf, this correspondence is 
obeyed in all the considered  nuclei. The MFPESs obtained for 
$^{296}$No (see Fig. \ref{Q2Q3_MF_No}) and $^{298}$Rf, exhibit 
a pronounced competition between two minima at almost the 
same energy but with quite different 
$Q_{20}$ and $Q_{30}$ deformations. This shape 
coexistence  leads to a rather low excitation energy of the 
$1^{-}$ state. However, the collective wave functions 
for the ground and negative parity states 
barely overlap, leading to a reduction 
in the  B(E3) value. For other nuclei, the ground 
state collective wave function is peaked at a non-zero octupole 
deformation and therefore strongly overlaps with the one
 of the negative parity state 
 (see, Figs.~\ref{COLLWS_POS_PARITY_No} and \ref{COLLWS_NEG_PARITY_No})
 leading to a large B(E3) value. For less 
octupole correlated systems, the peak of the ground state collective wave 
function shifts to $Q_{30}=0$ and therefore the overlap with the 
negative parity collective wave function is severely reduced as it is 
the E3 strength. As can be seen from the figures, the 1D-GCM and 2D-GCM
B(E3)
transition probabilities display
the same trend with the most pronounced quantitative differences 
being obtained for U, Pu, Cm, Cf and Fm isotopes around N = 194.

In the comparison between  1D-GCM and 2D-GCM calculations, one should 
keep in mind that even when the corresponding collective wave functions 
look similar along the octupole direction, their tiny differences 
can be associated with the differences in the results. The 
comparison between 1D-GCM and 2D-GCM results
in Figs.~\ref{E1BE1BE3-GCM-1}  and \ref{E1BE1BE3-GCM-2} reveals 
that, to a large extent, there is a decoupling between the quadrupole 
and octupole degrees of freedom in the studied nuclei and indicates that
the 1D-GCM framework represents a valuable computational tool to account 
for the systematics of the 1$^{-}$ excitation energies and transition 
probabilities in this exotic region of the nuclear chart.

Finally, in order to illustrate the robustness of the 2D-GCM 
predictions  with respect to the underlying Gogny-EDF, calculations 
have also been carried out with the parametrizations D1S, D1M$^{*}$ and 
D1M$^{**}$ for $^{288-304}$No. The results are depicted in 
Fig.~\ref{E1BE1BE3-GCM-sensitivity}. The largest quantitative 
differences are obtained with the D1S parametrization, as expected, 
because D1M$^{*}$ and D1M$^{**}$ were fitted to be as close as possible 
to D1M. However, from the comparison we conclude that the predicted 
trends, with neutron number, of the 1$^{-}$ excitation energies and 
reduced transition probabilities are rather insensitive to the 
Gogny-EDF employed in the calculations.
 
\section{Conclusions} 
\label{conclusions} 

In this paper we have studied the interplay between quadrupole and 
octupole degrees of freedom in a set of  even-even neutron-rich 
actinides and superheavy nuclei with $92 \le$ Z $\le 110$ and  $186 
\le$ N $\le 202$ both at the mean-field level and beyond. To this end, 
we have resorted to the Gogny-HFB framework, parity projection and 
2D-GCM  configuration mixing calculations with the quadrupole $Q_{20}$ 
and octupole $Q_{30}$ moments as generating coordinates.

Static octupole deformations are found around the "octupole neutron 
magic number" N=192 in roughly 30 \% of the 90 nuclei analyzed. On the 
other hand, dynamical octupole deformations are ubiquitous and have a 
significant impact on correlation energies leading to a weaker 
dependence with neutron number. The consideration of beyond-mean-field 
effects within the 2D-GCM  approach allows us to explore properties of 
the lowest-lying collective negative parity excited states such as, 
their excitation energies and transition strengths to the positive 
parity ground state. Low excitation energies and large E3 strengths are 
observed in nuclei with strong octupole correlations. The E1 strength, 
as expected, is a less collective quantity and  does not show a clear 
correlation with octupole properties. 

Given that very neutron-rich isotopes are considered, a comparison with 
experimental data is not possible at present and probably will not be 
possible in the future. However, the properties analyzed can be used to 
model the nuclear reactions taking place in the r-process 
nucleo-synthesis of superheavy nuclei. The relevance of this lies on 
the fact that the population of short lived superheavy nuclei is 
thought to have an impact on the solar abundance of mid-mass elements 
in the rare earth region through fission recycling.

\begin{acknowledgments} 
The  work of LMR was supported by Spanish 
Ministry of Economy and Competitiveness (MINECO) Grants No. 
PGC2018-094583-B-I00. We acknowledge the computer resources and 
assistance provided by Centro de Computación Científica-Universidad 
Autónoma de Madrid (CCC-UAM). 
\end{acknowledgments}

\end{document}